\documentclass[aps,prd,reprint,twocolumn,10pt,eqsecnum,showpacs,nofootinbib,longbibliography,superscriptaddress,floatfix]{revtex4-2}
\usepackage[colorlinks=true,allcolors=blue]{hyperref}
\usepackage{bm}
\usepackage{color}
\usepackage{amsfonts}
\usepackage{amssymb}
\usepackage{amsmath}
\usepackage{mathrsfs}
\usepackage{mathtools}

\newcommand{\ud}{\mathrm{d}}
\newcommand{\uD}{\mathrm{D}}

\newcommand{\pbg}{\,\mbox{}_\gamma}
\newcommand{\bs}[1]{\boldsymbol{#1}}
\newcommand{\mc}[1]{\mathcal{#1}}
\newcommand{\ms}[1]{\mathscr{#1}}
\newcommand{\ord}[2]{\underset{^{(#1)}}{#2}{}}
\newcommand{\vkappa}[1]{\smash{\overset{_\varkappa}{#1}}}

\begin{document}

\author{Alexander M.\ Grant}
\email{a.m.grant@soton.ac.uk}
\affiliation{School of Mathematical Sciences, University of Southampton,
  Southampton, SO17 1BJ, United Kingdom}

\title{Persistent gravitational wave observables: Nonlinearities in (non-)geodesic deviation}

\begin{abstract}
  The usual gravitational wave memory effect can be understood as a change in the separation of two initially comoving observers due to a burst of gravitational waves.
  Over the past few decades, a wide variety of other, ``persistent'' observables which measure permanent effects on idealized detectors have been introduced, each probing distinct physical effects.
  These observables can be defined in (regions of) any spacetime where there exists a notion of radiation, such as perturbation theory off of a fixed background, nonlinear plane wave spacetimes, or asymptotically flat spacetimes.
  Many of the persistent observables defined in the literature have only been considered in asymptotically flat spacetimes, and the perturbative nature of such calculations has occasionally obscured deeper relationships between these observables that hold more generally.
  The goal of this paper is to show how these more general results arise, and to do so we focus on two observables related to the separation between two, potentially accelerated observers.
  The first is the \emph{curve deviation}, which is a natural generalization of the displacement memory, and also contains what this paper proposes to call \emph{drift} memory (previously called ``subleading displacement memory'') and \emph{ballistic} memory.
  The second is a relative proper time shift that arises between the two observers, either at second order in their initial separation and relative velocity, or in the presence of relative acceleration.
  The results of this paper are, where appropriate, entirely non-perturbative in the curvature of spacetime, and so could be used beyond leading order in asymptotically flat spacetimes.
\end{abstract}

\maketitle

\tableofcontents

\section{Introduction}

This paper continues the study of persistent gravitational wave observables introduced in~\cite{Flanagan2019a} (hereafter Paper~I), which was then further developed in~\cite{Flanagan2019c, Grant2021b} (hereafter Papers~II and~III).
Paper~I introduced these observables as a general class which encompasses the many generalizations of the gravitational\footnote{Here, and throughout the rest of this paper, we restrict the discussion to gravitational effects, although similar effects in a variety of other theories have been considered as well (see, for example,~\cite{Bessonov1981, Bieri2013, Pasterski2015b, Pate2017, Ferko2021, Garfinkle2022, Oblak2023, Perez2023, Sheikh-Jabbari2023}).} wave memory effect (see, for example,~\cite{Grishchuk1989, Strominger2014b, Flanagan2014, Pasterski2015a, Nichols2017, Nichols2018}) that had arisen since its discovery by Zel'dovich and Polnarev~\cite{Zeldovich1974}.
These observables were designed to be quite general, and were defined in any class of spacetimes in which a notion of radiation could be defined: these include
\begin{itemize}

\item perturbations off of a known, exact spacetime (considered for flat backgrounds in Paper~I and, for example,~\cite{Zeldovich1974, Grishchuk1989, Thorne1992, Bieri2014, Flanagan2014}, as well for cosmological backgrounds in, for example,~\cite{Bieri2015, Tolish2016, Jokela2022});

\item exact, nonlinear plane wave spacetimes (considered in Paper~II and, for example,~\cite{Harte2015, Zhang2017a, Zhang2017b}); and

\item asymptotically flat spacetimes (considered in Paper~III and, for example,~\cite{Christodoulou1991, Strominger2014a, Strominger2014b, Pasterski2015a, Nichols2017, Nichols2018}).

\end{itemize}

Explicitly, the definition of a persistent observable is the following: consider a set of observers in such a class of spacetimes.
A persistent observable is a set of measurements which these observers can perform, over some interval of their proper time, that will be nonzero only in the case where gravitational radiation had been present during this interval.\footnote{Note that this definition is the contrapositive of the original definition in Paper~I, and so is equivalent.}
For example, the original memory effect (now called the \emph{displacement} memory effect to distinguish it from other effects~\cite{Strominger2014b}) is a change (over some interval of time) in the separation of a pair of two initially comoving observers who follow geodesics.
Up to some caveats discussed in the next paragraph, this separation would be unchanged in the absence of radiation.

Note that the class of spacetimes in which one is considering an observable can significantly affect whether or not it counts as ``persistent''.
In Paper~I, for example, the most general class of spacetimes which were considered were those that contained a ``flat-to-flat'' transition: two regions through which the observers traveled where spacetime was locally flat, separated by a region that (potentially) possessed a non-zero curvature tensor.
The presence of non-zero curvature in this intermediate region represented the existence of radiation.
While this is an acceptable model for nonlinear plane wave spacetimes (and so was also used in Paper~II), it is \emph{only} a valid model for asymptotically flat spacetimes up to a certain order in $1/r$.
In asymptotically flat spacetimes, the natural notion of radiation is the non-vanishing of the \emph{Bondi news tensor}, and in regions where it does vanish, the curvature is $O(1/r^3)$, as it is in (say) the Schwarzschild spacetime.
As such, the displacement memory should only truly be a persistent observable, in a perturbative sense, up to $O(1/r^2)$.
This is expected, as even in the absence of radiation, the separation between two freely-falling observers will change due to tidal forces.

Since Paper~I, several new persistent observables have been defined~\cite{Chakraborty2020, Seraj2021a, Divakarla2021, Seraj2021b, Seraj2022a, Seraj2022b}.
Of interest to this paper is that many of these effects have been considered only in asymptotically flat spacetimes.
This has been the natural context, both due to practical concerns (as most gravitational wave phenomena which we observe arise for distant, isolated systems) and because there are nice interpretations of (some) persistent observables in terms of asymptotic symmetries of such spacetimes (see~\cite{Strominger2017} and the references therein).
However, by only considering observables at leading [$O(1/r)$, as in Paper~III], or at most the first subleading [$O(1/r^2)$] order, it is possible that more general features of these observables have been missed: in particular, those which are \emph{nonlinear} in the curvature tensor.\footnote{Since the distance to a source of gravitational waves is typically assumed to be large for astrophysically-relevant sources, considering $O(1/r^2)$ corrections arising from nonlinear effects may seem pointless, from a practical perspective.
  However, this is only true if the measurement is on a timescale which is sufficiently short compared to the period of the gravitational wave.
  Since the curvature scales as $\omega^2 h$, where $\omega$ is the frequency of the gravitational wave and $h$ its amplitude, if the measurement timescale $T \sim 1/(\omega \sqrt h)$, either because $T$ \emph{or} $\omega$ is large, then one cannot assume that nonlinear-in-curvature effects are small.
  For this reason, one should be careful when considering the effect of gravitational waves on test particles in the limit of infinite time.}
In contrast, for nonlinear plane wave spacetimes, one can obtain truly nonlinear results (as was done in Paper~II).
However, nonlinear plane wave spacetimes are not a useful model for studying radiation from isolated sources on scales where the spherical nature of the gravitational waves becomes apparent.

The main goal of this paper is to return to the original spirit of Paper~I by considering features of persistent observables in more general spacetimes, beyond the classes of spacetimes in which these features are typically discussed.
In particular, we are interested in relationships both between different persistent observables and between persistent observables and more fundamental, observable quantities, in contexts where the curvature is not assumed to be weak.

One motivation for doing this is to provide alternative ways to effectively measure an observable.
For example, the original spin memory observable, as defined by~\cite{Pasterski2015a}, was defined in terms of a phase shift appearing between two counterrotating beams of light in a Sagnac interferometer.
In asymptotically flat spacetimes, it was shown that this effect is $O(1/r^2)$, which is far too small to measure for astrophysical sources.
In contrast, a different procedure was proposed by~\cite{Nichols2017} (also bearing the name ``spin memory'') which gave an equivalent expression, but at $O(1/r)$, which is more feasible to observe.
In a similar vein, current proposals (for example~\cite{Lasky2016, Grant2022c}) for measuring persistent observables do not consider setting up the appropriate system of observers, but instead suggest taking data from a gravitational wave detector and (up to technical caveats, see~\cite{Grant2022c} for a discussion) showing that the persistent observable would have been non-zero if appropriate observers had been present.
As such, it is important to know how these observables are related to one another (to determine which observables a given measurement can probe) and to more easily-measured quantities, such as the curvature along an observer's worldline as a function of time.

The first nonlinear relationship which we consider is one that was noted in Paper~II, for exact plane wave spacetimes.
Here, many of the persistent observables which had been defined in Paper~I, while describing many disparate physical effects, could all be written in terms of a set of four ($2 \times 2$-matrix-valued) functions of the coordinates.
These four functions, which were called in Paper~II the \emph{transverse Jacobi propagators} (as they were the in-plane components of the Jacobi propagators discussed in Sec.~\ref{sec:jacobi}), could in fact be used to solve the geodesic equation in plane wave spacetimes \emph{exactly}, and so arose naturally for any observables that were defined in terms of geodesics.
Moreover, these transverse Jacobi propagators had the property that they were not all independent: their values were constrained by the existence of a conserved Wro\'nskian.
It was argued in Paper~II that this would mean that not all observables related to geodesic motion would be independent in these spacetimes.
Moreover, at leading order in $1/r$ in asymptotically flat spacetimes, this result has already been known~\cite{Seraj2021a, Divakarla2021}.
In Sec.~\ref{sec:redundant}, we show explicitly that this result holds more generally, in particular at subleading orders in $1/r$.

The second relationship which we consider is one that relates persistent observables and integrated notions of radiation.
In Paper~I, it was noticed that the persistent  observables which were considered could be characterized entirely by a set of temporal moments of the curvature tensor along an observer's worldline, to first order in the curvature.
This was used in Paper~III to prove a similar result asymptotically for the curve deviation observable from Paper~I, to leading order in $1/r$, in terms of temporal moments of the Bondi news tensor.
This was, perhaps, to be expected: when considering effects which are nonlocal in time and are nonzero only if there is radiation during the interval of time in question, the only expressions which would naturally arise would be integrals of the quantity which characterizes the presence of radiation.
In the case of transitions between flat regions, the curvature tensor characterizes the presence of radiation, while in transitions between nonradiative regions in asymptotically flat spacetimes, this role is played by the Bondi news.
As we will show in Sec.~\ref{sec:moments}, there exists a natural generalization of these results: the curve deviation can generally be written in terms of the moments of an appropriately-defined bitensor.

Throughout this paper, we focus entirely on observables which arise from two observers, and are related to measurements of their relative separation.
Moreover, most of the observables which we consider, while nonlinear in curvature, are linear in the separation, relative velocity, and relative acceleration of these two observers.
However, as a bonus, we also consider in Sec.~\ref{sec:proper_time} a proper time shift observable which was first discovered for observers in asymptotically flat spacetimes~\cite{Strominger2014b} which, for initially comoving, freely falling observers, is \emph{nonlinear} in the initial separation.
While this observable has been considered earlier in this series in the contexts of linearized gravity and nonlinear plane wave spacetimes (in Papers~I and~II, respectively), we present the result for more general spacetimes for the first time.

The structure of the remainder of this paper is as follows.
First, in Sec.~\ref{sec:bitensors}, we set the stage for computing the persistent observables which appear in this paper, which (as mentioned above) are constructed entirely from the separation of two closely-separated observers.
We do so by using the covariant theory of bitensors, which we review in this section, and we also review the derivation of formulas for the evolution of the separation in terms of these bitensors, generalizing the usual formulae for nonlinear geodesic deviation (as derived in, say,~\cite{Vines2014a}) to the non-geodesic case.
We also explicitly discuss the Wro\'nskian relationship for a set of bitensors known as the Jacobi propagators, which (as mentioned above) has applications to determining relationships between certain persistent observables.
Next, in Sec.~\ref{sec:persistent_observables}, we present our results for the curve deviation observable, both in terms of nonlinear relationships between its various pieces and a generalization of the ``moments'' which were used in Papers~I and~III to understand this observable.
We also present our results for the proper time shift in this same section.
Finally, we conclude with a discussion in Sec.~\ref{sec:discussion}.

The notation in this paper is as follows: we adopt the mostly plus metric signature, and follow the conventions for the curvature tensor of Wald~\cite{Wald1984}.
Apart from a few exceptions, we use only abstract indices in this paper, which we denote using Latin letters from the beginning of the alphabet ($a$, $b$, etc.).
We use the conventions for bitensors from~\cite{Poisson2011}, and following Paper~I we only explicitly give the dependence of a bitensor at a point when it is a scalar at that point, and use the same annotations for points and their corresponding indices [for example, $\omega_{\bar a} (x')$ denotes a one-form at $\bar x$ and a scalar at $x'$].
Moreover, for any curve $\gamma$, we denote by $x$ the point $\gamma(\tau)$, and apply any adornments to $\tau$ or $\gamma$ to $x$ as well [so, for example, $\bar x' \equiv \bar \gamma(\tau')$].
Finally, we will frequently take products of order symbols, so that (for example) $O(a, b)^2$ is shorthand for $O(a^2, ab, b^2)$.

\section{Review of covariant bitensors} \label{sec:bitensors}

Following Paper~I (and to a lesser extent Paper~II, and to an even lesser extent, Paper~III), we study persistent observables in this paper using the theory of covariant bitensors.
These are tensor fields which are a function of two spacetime points, which moreover can have a tensorial character at each of these points.
As bitensors do not typically arise in discussions of gravitational wave memory, we review their properties in the sections below.

The relevance of bitensors to the gravitational wave memory effect is twofold.
First, bitensors can be used to prove (generalizations of) the geodesic deviation equation, as shown in~\cite{Aleksandrov1979, Vines2014a}.
As the geodesic deviation equation and its generalizations form the foundation of the gravitational wave memory effect, we review this derivation in detail below, culminating in Eq.~\eqref{eqn:non_geodesic}.
Bitensors are also the natural language for studying persistent observables more generally, as they provide a coordinate-independent way to describe nonlocal quantities, such as the difference of two tensors evaluated at different points or solutions of differential equations along a curve.

In particular, two bitensors, the \emph{Jacobi propagators} $\pbg K^{a'}{}_a$ and $\pbg H^{a'}{}_a$, will be key to the discussion of the rest of the paper.
As we will show in Sec.~\ref{sec:jacobi}, they can be used to construct the solution to the geodesic deviation equation along a curve $\gamma$:
\begin{equation}
  \xi^{a'} = \pbg K^{a'}{}_a \xi^a + (\tau' - \tau) \pbg H^{a'}{}_a \dot \xi^a + O(\bs \xi, \dot{\bs \xi})^2.
\end{equation}
Here, $\xi^{a'}$ is the separation vector at some time $\tau'$ along $\gamma$, and $\xi^a$ and $\dot \xi^a$ are the separation and relative velocity at the initial time $\tau$, respectively.
As we will describe in more detail in Sec.~\ref{sec:curve_dev}, $\pbg K^{a'}{}_a$ and $(\tau' - \tau) \pbg H^{a'}{}_a$ give rise to the displacement and drift (that is, ``subleading displacement'') memories, respectively.

\subsection{Fundamental bitensors}

In this section, we review two bitensors which lay the foundation for the discussion in this paper, namely Synge's world function and the parallel propagator.
Their primary utility is in the fact that one can use them to study the Taylor expansion of any bitensor, as a function of a separation vector defined between the two points at which the bitensor is evaluated.
Naturally arising in this discussion is the notion of a coincidence limit, the limit where these two points are taken to be the same.

We start with Synge's world function, which is a scalar function $\sigma(x, x')$ of two points $x$ and $x'$, defined to be half of the square of the geodesic distance between these two points.
This bitensor is only defined for points which lie within a convex normal neighborhood of one another, which means that there is a unique geodesic which connects these two points.
As this quantity is a function of two points, one can take (potentially repeated) derivatives with respect to either point, which commute: as is conventional, we denote these by simply appending indices to the end of $\sigma$:
\begin{equation}
  \sigma_{a_1 \cdots a_r b_1' \cdots b_s'} \equiv \nabla_{a_r} \cdots \nabla_{a_1} \nabla_{b_s'} \cdots \nabla_{b_1'} \sigma(x, x').
\end{equation}
The most important of these derivatives are $-\sigma_a (x')$ and $\sigma_{a'} (x)$, as one can show that (when raised with the metric) they are the tangent vectors to the unique geodesic between $x$ and $x'$, where this geodesic is parameterized such that $x$ is where the parameter is zero and $x'$ is where the parameter is one.
As such, $-\sigma^a (x')$ has a natural interpretation as a ``separation vector'' between $x$ and $x'$.
Moreover, it follows that
\begin{equation}
  \sigma(x, x') = \frac{1}{2} \sigma^a (x') \sigma_a (x') = \frac{1}{2} \sigma^{a'} (x) \sigma_{a'} (x),
\end{equation}
and differentiating these expressions with respect to $x$ or $x'$ gives
\begin{equation} \label{eqn:synge_geodesic}
  \sigma^b (x') \sigma^a{}_b (x') = \sigma^a (x') = \sigma^{b'} (x') \sigma^a{}_{b'},
\end{equation}
which is just the non-affinely parameterized geodesic equation.

Using these separation vectors, we can expand bitensors in the following way (see, for example,~\cite{Vines2014b, Poisson2011}): for any bitensor of the form $\Omega_{\ms A} (x')$ whose indices, which we denote with a composite index $\ms A$ \{see, for example, the comments below Proposition~(2.2.35) of~\cite{Penrose1987}\}, lie entirely at $x$, we seek an expansion of the form
\begin{equation} \label{eqn:coincident_expansion}
  \Omega_{\ms A} (x') = \sum_{n = 0}^\infty \frac{(-1)^n}{n!} \Omega_{\ms A b_1 \cdots b_n} \sigma^{b_1} (x') \cdots \sigma^{b_n} (x'),
\end{equation}
where each term in this expansion is, most importantly, \emph{only} a tensor at $x$.
Here, loosely following~\cite{Vines2014b}, we will derive a version of Taylor's theorem, determining the coefficients of this expansion through repeated differentiation.

To do so, we first need to define the \emph{coincidence limit} of a bitensorial expression, which corresponds to the limit of taking the two points $x$ and $x'$ to coincide.
This is simple in the case where we consider a bitensor of the form $\Omega_{\ms A} (x')$, where all indices lie at $x$: the definition is given in terms of the usual limit:
\begin{equation}
  [\Omega_{\ms A} (x')]_{x' \to x} \equiv \lim_{x' \to x} \Omega_{\ms A} (x').
\end{equation}
In the case where there are some indices (let us denote them by a composite index $\ms B'$) which lie at $x'$, the coincidence limit is defined by
\begin{equation}
  [\Omega_{\ms A \ms B'}]_{x' \to x} \equiv \lim_{x' \to x} (\Omega_{\ms A \ms B'} Z^{\ms B'}{}_{\ms B}),
\end{equation}
where $Z^{\ms A'}{}_{\ms A}$ is any bitensor such that $Z^{\ms A}{}_{\ms B} = \delta^{\ms A}{}_{\ms B}$, the identity.
Note that this is independent of which $Z^{\ms A'}{}_{\ms A}$ one chooses, so long as this limit is well-defined.
Furthermore, coincidence limits obey an identity known as \emph{Synge's rule}~\cite{Synge1960, Poisson2011}:
\begin{equation} \label{eqn:synges_rule}
  \nabla_a [\Omega_{\ms B \ms C'}]_{x' \to x} = [\nabla_a \Omega_{\ms B \ms C'}]_{x' \to x} + [\nabla_{a'} \Omega_{\ms B \ms C'}]_{x' \to x}.
\end{equation}

A variety of coincidence limits are well-known: in particular, it is clear that
\begin{gather}
  [\sigma(x, x')]_{x' \to x} = 0, \\
  [\sigma_a (x')]_{x' \to x} = [\sigma_{a'} (x)]_{x' \to x} = 0. \label{eqn:sep_coincidence}
\end{gather}
Moreover, by using Eqs.~\eqref{eqn:synge_geodesic} and~\eqref{eqn:synges_rule}, one can show that~\cite{Synge1960, Poisson2011}
\begin{equation} \label{eqn:synge_coincidence}
  \begin{split}
    \delta^a{}_b &= [\sigma^a{}_b (x')]_{x' \to x} = [\sigma^{a'}{}_{b'} (x)]_{x' \to x}  \\
    &= -[\sigma^a{}_{b'}]_{x' \to x} = -[\sigma^{a'}{}_b]_{x' \to x}.
  \end{split}
\end{equation}

As such, upon taking $k$ symmetrized derivatives of Eq.~\eqref{eqn:coincident_expansion} with respect to $x'$, three cases arise once one takes the coincident limit.
For $k < n$, the coincidence limit vanishes, as there will be remaining factors of $\sigma^{b_i} (x')$ remaining.
For $k = n$, the only contribution to the coincidence limit is $\Omega_{\ms A b_1 \cdots b_n}$.
However, if $k > n$, there will be terms which involve coincidence limits of symmetrized derivatives of $\sigma^a (x')$.
As we show in Appendix~\ref{app:symmetrized}, these coincidence limits vanish, and so
\begin{equation} \label{eqn:scalar_taylor}
  \begin{split}
    \Omega_{\ms A} (x') = \sum_{n = 0}^\infty \frac{(-1)^n}{n!} &\sigma^{b_1} (x') \cdots \sigma^{b_n} (x') \\
    \times &[\nabla_{(b_1'} \cdots \nabla_{b_n')} \Omega_{\ms A} (x')]_{x' \to x}.
  \end{split}
\end{equation}
In the case where we have a bitensor of the form $\Omega_{\ms A \ms B'}$, we can instead perform the expansion for $\Omega_{\ms A \ms B'} Z^{\ms B'}{}_{\ms B}$, and then apply $(Z^{-1})^{\ms B}{}_{\ms C'}$.
Here, unlike before, the choice of $Z^{\ms A'}{}_{\ms A}$ is important, as the expansion involves taking derivatives of this bitensor.

We finally turn to a common choice of such a bitensor $Z^{a'}{}_a$, the \emph{parallel propagator} $\pbg g^{a'}{}_a$.
Given a curve $\gamma$, consider a basis $(e_\alpha)^a$ at some point $x$ along $\gamma$, together with its dual basis $(\omega^\alpha)_a$.
Then, extend these bases to be functions along $\gamma$ by parallel transport:
\begin{equation}
  \dot \gamma^b \nabla_b (e_\alpha)^a = 0, \qquad \dot \gamma^b \nabla_b (\omega^\alpha)_a = 0.
\end{equation}
Note that $\nabla_a$ here can be \emph{any} connection, not necessarily the metric-compatible one (although in this paper, it typically will be).
The parallel propagator is then defined by~\cite{Poisson2011}
\begin{equation}
  \pbg g^{a'}{}_a = \sum_\alpha (e_\alpha)^{a'} (\omega^\alpha)_a.
\end{equation}
Note that this parallel propagator is always defined in terms of a given curve, which we indicate with the initial $\gamma$ subscript.
However, it is occasionally useful to consider the parallel propagator defined with respect to the unique geodesic connecting two points in a convex normal neighborhood: we denote this by $g^{a'}{}_a$.

The important properties of the parallel propagator are as follows, and can be readily verified from its definition:
\begin{subequations} \label{eqn:parallel_composition}
  \begin{align}
    \pbg g^{a''}{}_{a'} \pbg g^{a'}{}_a &= \pbg g^{a''}{}_a, \\
    \pbg g^a{}_{a'} \pbg g^{a'}{}_b &= \delta^a{}_b,
  \end{align}
\end{subequations}
along with the following two properties, which, in a sense, provide an abstract definition:
\begin{equation} \label{eqn:parallel_def}
  \dot \gamma^{b'} \nabla_{b'} \pbg g^{a'}{}_a = \dot \gamma^b \nabla_b \pbg g^{a'}{}_a = 0.
\end{equation}
Moreover, if $\gamma$ is a geodesic, it follows from the geodesic equation that
\begin{equation}
  \pbg g^{a'}{}_a \dot \gamma^a = \dot \gamma^{a'}, \qquad \dot \gamma_{a'} \pbg g^{a'}{}_a = \dot \gamma_a.
\end{equation}

In terms of Taylor's theorem, the parallel propagator is particularly convenient to use, as symmetrized derivatives of $g^{a'}{}_a$ vanish under the coincidence limit, as shown in Appendix~\ref{app:symmetrized}.
As such, it follows that
\begin{equation} \label{eqn:general_taylor}
  \begin{split}
    \Omega_{\ms A \ms B'} = \sum_{n = 0}^\infty \frac{(-1)^n}{n!} &\sigma^{c_1} (x') \cdots \sigma^{c_n} (x') \\
    \times &g^{\ms D}{}_{\ms B'} [\nabla_{(c_1'} \cdots \nabla_{c_n')} \Omega_{\ms A \ms D'}]_{x' \to x},
  \end{split}
\end{equation}
where $g^{\ms A}{}_{\ms A'}$ acts on each index in $\ms A$ through a parallel propagator (for example, if $\ms A$ contains two raised indices $a$ and $b$ and a lowered index $c$, then $g^{ab}{}_{ca'b'}{}^{c'} = g^a{}_{a'} g^b{}_{b'} g^{c'}{}_c$).

\subsection{Jacobi propagators} \label{sec:jacobi}

We now turn to the most important bitensors in this section, the \emph{Jacobi propagators}.
Locally, these can be defined in terms of Synge's world function by~\cite{Dixon1979}
\begin{equation} \label{eqn:jacobi_def}
  H^{a'}{}_a \equiv -(\sigma^{-1})^{a'}{}_a, \qquad K^{a'}{}_a \equiv H^{a'}{}_b \sigma^b{}_a (x').
\end{equation}
The former of these expressions has the following meaning:
\begin{equation} \label{eqn:sigma_inverse}
  H^{a'}{}_a \sigma^a{}_{b'} = -\delta^{a'}{}_{b'};
\end{equation}
as can be readily verified, this is equivalent to
\begin{equation}
  \sigma^a{}_{a'} H^{a'}{}_b = -\delta^a{}_b.
\end{equation}

By differentiating Eq.~\eqref{eqn:sigma_inverse}, we can recover
\begin{subequations}
  \begin{align}
    \nabla_{c'} H^{a'}{}_b &= H^{a'}{}_a H^{b'}{}_b \sigma^a{}_{b'c'}, \\
    \nabla_c H^{a'}{}_b &= H^{a'}{}_a H^{b'}{}_b \sigma^a{}_{b'c}.
  \end{align}
\end{subequations}
Similarly, taking a derivative of $K^{a'}{}_a$ using Eq.~\eqref{eqn:jacobi_def} gives
\begin{subequations}
  \begin{align}
    \nabla_{c'} K^{a'}{}_b &= H^{a'}{}_a (K^{b'}{}_b \sigma^a{}_{b'c'} + \sigma^a{}_{bc'}), \\
    \nabla_c K^{a'}{}_b &= H^{a'}{}_a [K^{b'}{}_b \sigma^a{}_{b'c} + \sigma^a{}_{bc} (x')].
  \end{align}
\end{subequations}
Moreover, by taking derivatives of Eq.~\eqref{eqn:synge_geodesic} (together with the version with $x$ and $x'$ flipped) we can eliminate the third derivatives of Synge's world function from these equations by contracting with $\sigma^{c'} (x)$ and $\sigma^c (x')$, yielding the following expressions:
\begin{subequations}
  \begin{align}
    \sigma^{c'} (x) \nabla_{c'} H^{a'}{}_b &= -H^{a'}{}_b + g_{bc} K^c{}_{c'} g^{c'a'}, \label{eqn:prime_dH} \\
    \sigma^c (x') \nabla_c H^{a'}{}_b &= -H^{a'}{}_b + K^{a'}{}_b, \label{eqn:dH} \\
    H^a{}_{b'} \sigma^{c'} (x) \nabla_{c'} K^{b'}{}_b &=  K^a{}_{b'} K^{b'}{}_b - \delta^a{}_b, \label{eqn:prime_dK} \\
    \sigma^c (x') \nabla_c K^{a'}{}_b &= -H^{a'}{}_a R^a{}_{cbd} \sigma^c (x') \sigma^d (x'). \label{eqn:dK}
  \end{align}
\end{subequations}

At this point, affinely parameterize the geodesic $\gamma$ between $x$ and $x'$, and let $\tau$ be the value of this affine parameter at $x$ and $\tau'$ its value at $x'$ (in many of the cases in this paper, $\gamma$ will be timelike and $\tau$ proper time, but this is not necessary).
One can then show starting from these equations that $K^{a'}{}_a$ and $(\tau' - \tau) H^{a'}{}_a$ are both solutions to the following differential equation:
\begin{equation} \label{eqn:jacobi}
  \frac{\uD^2}{\ud \tau'^2} A^{a'}{}_a = -R^{a'}{}_{\dot \gamma' b' \dot \gamma'} A^{b'}{}_a,
\end{equation}
where we have defined
\begin{equation} \label{eqn:dotgamma_contract}
  R^a{}_{\dot \gamma b \dot \gamma} \equiv R^a{}_{cbd} \dot \gamma^c \dot \gamma^d.
\end{equation}
Note that, upon contracting Eq.~\eqref{eqn:jacobi} with any vector $v^a$ at $x$, one recovers the (leading-order) geodesic deviation equation for $\xi^{a'} \equiv A^{a'}{}_a v^a$:
\begin{equation}
  \ddot \xi^{a'} = -R^{a'}{}_{\dot \gamma' b' \dot \gamma'} \xi^{b'}.
\end{equation}
It is for this reason that the Jacobi propagators are useful for solving the geodesic deviation equation.

These solutions to Eq.~\eqref{eqn:jacobi}, $K^{a'}{}_a$ and $(\tau' - \tau) H^{a'}{}_a$, differ only in their boundary conditions: by the coincidence limits of derivatives of Synge's world function, it is clear that
\begin{equation} \label{eqn:KH_init}
  [K^{a'}{}_b]_{x' \to x} = \delta^a{}_b, \quad [(\tau' - \tau) H^{a'}{}_b]_{x' \to x} = 0,
\end{equation}
and by applying Eqs.~\eqref{eqn:prime_dH} and~\eqref{eqn:prime_dK} and coincidence limits, we find that
\begin{equation} \label{eqn:dKH_init}
  \left[\frac{\uD K^{a'}{}_b}{\ud \tau'}\right]_{x' \to x} = 0, \quad \left[\frac{\uD \{(\tau' - \tau) H^{a'}{}_b\}}{\ud \tau'}\right]_{x' \to x} = \delta^a{}_b.
\end{equation}
Under appropriate differentiability conditions on the metric, these are the unique solutions to Eq.~\eqref{eqn:jacobi} satisfying these boundary conditions, as Eq.~\eqref{eqn:jacobi} is simply an ordinary differential equation along $\gamma$.

As with the parallel propagator, one can consider the differential equation in Eq.~\eqref{eqn:jacobi} along \emph{any} curve, and then use that in order to define curve-dependent Jacobi propagators $\pbg K^{a'}{}_a$ and $\pbg H^{a'}{}_a$.
We discuss the properties of these more general Jacobi propagators in the next two sections.
This discussion is primarily based upon that in Sec.~4.2.1.3 of~\cite{GrantThesis}.

\subsubsection{Definition for any curve}

First, we note that the analogy between the parallel and Jacobi propagators runs quite deep: first, consider the space of vector fields $X^A$ defined by
\begin{equation}
  X^A \equiv \begin{pmatrix}
    \xi^a \\
    \dot \xi^a
  \end{pmatrix};
\end{equation}
this vector field can be considered as a section on the direct-, or Whitney-sum bundle of two copies of the tangent bundle to the manifold.
Equivalently, at each point it is a member of the direct sum of two copies of the tangent space.
Here, we use capital Latin letters from the beginning of the alphabet for the indices on this larger space, following~\cite{Flanagan2016} and Paper~I.
One can then consider a bitensor $\pbg J^{A'}{}_A$ that is defined by
\begin{equation} \label{eqn:jacobi_matrix}
  \pbg J^{A'}{}_A \equiv \begin{pmatrix}
    \pbg K^{a'}{}_a & (\tau' - \tau) \pbg H^{a'}{}_a \\[1em]
    \dfrac{\uD \pbg K^{a'}{}_a}{\ud \tau'} & \dfrac{\uD \left[(\tau' - \tau) \pbg H^{a'}{}_a\right]}{\ud \tau'}
  \end{pmatrix}.
\end{equation}
One can further define a curve-dependent connection by
\begin{equation}
  \pbg \hat \nabla_c X^A = \nabla_c X^A + \hat C^A{}_{Bc} X^B,
\end{equation}
where $\nabla_a$ acts on each component of $X^A$ in the usual sense, and
\begin{equation}
  \pbg \hat C^A{}_{Bc} = \begin{pmatrix}
     0 & \dot \gamma_c \delta^a{}_b \\
     R^a{}_{cb \dot \gamma} & 0
  \end{pmatrix}.
\end{equation}
It is then the case that $\pbg J^{A'}{}_A$ \emph{is} the parallel propagator with respect to this connection, as
\begin{equation}
  \dot \gamma^{b'} \pbg \hat \nabla_{b'} \pbg J^{A'}{}_A = 0.
\end{equation}
In particular, this means that one can immediately derive the following expressions from Eqs.~\eqref{eqn:parallel_composition} and~\eqref{eqn:parallel_def}:
\begin{subequations}
  \begin{align}
    \pbg J^{A''}{}_{A'} \pbg J^{A'}{}_A &= \pbg J^{A''}{}_A, \label{eqn:jacobi_composition} \\
    \pbg J^A{}_{A'} \pbg J^{A'}{}_B &= \delta^A{}_B,
  \end{align}
\end{subequations}
together with
\begin{equation} \label{eqn:jacobi_init_deriv}
  \dot \gamma^b \pbg \hat \nabla_b \pbg J^{A'}{}_A = 0,
\end{equation}
where (as is usually done with connections) we have extended $\pbg \hat \nabla_a$ to act on covectors such that the Leibniz rule holds.

The next property which we consider is the fact that any solution to Eq.~\eqref{eqn:jacobi}, along a general curve, must be a linear combination of the two solutions $\pbg K^{a'}{}_a$ and $(\tau' - \tau) \pbg H^{a'}{}_a$.
In particular, $\uD \pbg K^{a'}{}_a/\ud \tau$  and $\uD [(\tau' - \tau) \pbg H^{a'}{}_a]/\ud \tau$ are also solutions to this equation (note that the derivatives act on $\tau$, not $\tau'$).
By using Synge's rule, it follows that
\begin{subequations}
  \begin{align}
    \left[\frac{\uD \pbg K^{a'}{}_b}{\ud \tau}\right]_{\tau' \to \tau} &= 0, \\
    \left[\frac{\uD \{(\tau' - \tau) \pbg H^{a'}{}_b\}}{\ud \tau}\right]_{\tau' \to \tau} &= -\delta^a{}_b,
  \end{align}
\end{subequations}
and, applying a second derivative, that
\begin{subequations}
  \begin{align}
    \left[\frac{\uD^2 \pbg K^{a'}{}_b}{\ud \tau' \ud \tau}\right]_{\tau' \to \tau} = R^a{}_{\dot \gamma b \dot \gamma}, \\
    \left[\frac{\uD^2 \{(\tau' - \tau) \pbg H^{a'}{}_b\}}{\ud \tau' \ud \tau}\right]_{\tau' \to \tau} = 0.
  \end{align}
\end{subequations}
In order to match these boundary conditions, we therefore find that
\begin{subequations}
  \begin{align}
    \frac{\uD \pbg K^{a'}{}_a}{\ud \tau} &= (\tau' - \tau) \pbg H^{a'}{}_b R^b{}_{\dot \gamma a \dot \gamma}, \label{eqn:dK_initial} \\
    \frac{\uD [(\tau' - \tau) \pbg H^{a'}{}_a]}{\ud \tau} &= -\pbg K^{a'}{}_a \label{eqn:dH_initial}
  \end{align}
\end{subequations}
 (see~\cite{Harte2012a} for the specialization of these formulae to the transverse Jacobi propagators).
In the geodesic case, a direct comparison with Eqs.~\eqref{eqn:dK} and~\eqref{eqn:dH} also verifies that these equations hold.

A final useful property of Jacobi propagators is that, for a geodesic $\gamma$, it follows upon contracting Eq.~\eqref{eqn:jacobi} with $\dot \gamma^a$ that
\begin{equation}
  \frac{\uD^2 (\pbg A^{a'}{}_a \dot \gamma^a)}{\ud \tau'^2} = 0,
\end{equation}
for $\pbg A^{a'}{}_a = K^{a'}{}_a$ or $(\tau' - \tau) H^{a'}{}_a$.
As such, we can write
\begin{equation}
  \pbg A^{a'}{}_a \dot \gamma^a = \pbg g^{a'}{}_a B^a + (\tau' - \tau) \pbg g^{a'}{}_a C^a,
\end{equation}
and a direct comparison with the initial conditions for $\pbg A^{a'}{}_a \dot \gamma^a$ shows that
\begin{equation}
  \pbg K^{a'}{}_a \dot \gamma^a = \pbg H^{a'}{}_a \dot \gamma^a = \dot \gamma^{a'}.
\end{equation}
A similar analysis shows that
\begin{equation} \label{eqn:jacobi_final_u}
  \dot \gamma_{a'} \pbg K^{a'}{}_a = \dot \gamma_{a'} \pbg H^{a'}{}_a = \dot \gamma_a.
\end{equation}

\subsubsection{Wro\'nskian relationships}

We now discuss the existence of a conserved Wro\'nskian for Eq.~\eqref{eqn:jacobi} (see, for example,~\cite{Uzun2018, Grasso2018} for discussion in general spacetimes, or~\cite{Harte2012a, Harte2012b} for the special case of plane-wave spacetimes).
Given two solutions $A^{a'}{}_a$ and $B^{a'}{}_a$ to Eq.~\eqref{eqn:jacobi}, define
\begin{equation}
  W_{ab} (\tau') \equiv g_{a'b'} \left(B^{b'}{}_b \frac{\ud}{\ud \tau'} A^{a'}{}_a - A^{a'}{}_a \frac{\ud}{\ud \tau'} B^{b'}{}_b\right).
\end{equation}
Note that $W_{ab} (\tau')$ is in fact independent of $\tau'$, as
\begin{equation}
  \begin{split}
    \frac{\ud}{\ud \tau'} W_{ab} (\tau') &= R_{a' \dot \gamma' c' \dot \gamma'} A^{c'}{}_a B^{a'}{}_b - R_{a' \dot \gamma' c' \dot \gamma'} A^{a'}{}_a B^{c'}{}_b \\
    &= 0,
  \end{split}
\end{equation}
using Eq.~\eqref{eqn:jacobi}, the fact that $R_{abcd} = R_{cdab}$, and a relabeling.
Considering Eq.~\eqref{eqn:jacobi} as a second order differential equation, $W_{ab} (\tau') \equiv W_{ab}$ is the conserved Wro\'nskian defined for these two solutions.

The simplest case to consider is where
\begin{equation}
  A^{a'}{}_a = (\tau' - \tau) \pbg H^{a'}{}_a, \qquad B^{a'}{}_a = \pbg K^{a'}{}_a.
\end{equation}
By the initial conditions in Eqs.~\eqref{eqn:KH_init} and~\eqref{eqn:dKH_init}, it follows that $W_{ab} = g_{ab}$, so that
\begin{equation} \label{eqn:KH_wronskian}
  \begin{split}
    g_{ab} = g_{a'b'} \Bigg\{&\pbg K^{b'}{}_b \frac{\uD [(\tau' - \tau) \pbg H^{a'}{}_a]}{\ud \tau'} \\
    &- (\tau' - \tau) \pbg H^{a'}{}_a \frac{\uD \pbg K^{b'}{}_b}{\ud \tau'}\Bigg\}.
  \end{split}
\end{equation}
It therefore follows that not all of the components of $\pbg J^{A'}{}_A$ are independent.
This was already apparent from Eqs.~\eqref{eqn:dK_initial} and~\eqref{eqn:dH_initial}, but the importance of this equation is that this is a \emph{pointwise} relationship, and not just a relationship showing that some of these components can be determined in terms of others by solving ordinary differential equations.

There are further results which follow from the conservation of $W_{ab}$, which can be used to relate Jacobi propagators with flipped arguments.
First, consider the case where
\begin{equation}
  A^{a'}{}_a = (\tau' - \tau) \pbg H^{a'}{}_a, \quad B^{a'}{}_a = (\tau' - \tau'') \pbg H^{a'}{}_{a''} \pbg g^{a''}{}_a.
\end{equation}
Evaluating $W_{ab}$ at $\tau' = \tau''$, we find that
\begin{equation}
  W_{ab} = -(\tau'' - \tau) g_{a''b''} \pbg H^{a''}{}_a \pbg g^{b''}{}_b,
\end{equation}
whereas at $\tau' = \tau$, we find
\begin{equation}
  W_{ab} = (\tau - \tau'') g_{ac} \pbg H^c{}_{b''} \pbg g^{b''}{}_b.
\end{equation}
Equating these expressions (and renaming $\tau''$ to $\tau'$), it follows that
\begin{equation} \label{eqn:H_swap}
  \pbg H^{a'}{}_a = g^{a'b'} g_{ab} \pbg H^b{}_{b'},
\end{equation}
which is equivalent to Etherington's reciprocity law~\cite{Etherington1933, Harte2012b, Perlick2004}.
Similarly, if one uses
\begin{equation}
  A^{a'}{}_a = \pbg K^{a'}{}_a, \qquad B^{a'}{}_a = \pbg K^{a'}{}_{a''} \pbg g^{a''}{}_a,
\end{equation}
we find that (at $\tau' = \tau''$)
\begin{equation}
  W_{ab} = g_{a''b''} \pbg g^{b''}{}_b \frac{\uD \pbg K^{a''}{}_a}{\ud \tau''}
\end{equation}
and (at $\tau' = \tau$)
\begin{equation}
  W_{ab} = -g_{ac} \frac{\uD \pbg K^c{}_{b''}}{\ud \tau} \pbg g^{b''}{}_b.
\end{equation}
It therefore follows that\footnote{Note that there is a typo in Eq.~(4.88) of~\cite{GrantThesis}: the derivative on the right-hand side of that equation should be with respect to $\tau$, not $\tau'$.}
\begin{equation} \label{eqn:K_swap}
  \frac{\uD \pbg K^{a'}{}_a}{\ud \tau'} = -g_{ab} g^{a'b'} \frac{\uD \pbg K^a{}_{a'}}{\ud \tau}.
\end{equation}

These equations allow us to derive the equivalents of Eqs.~\eqref{eqn:prime_dH} and~\eqref{eqn:prime_dK}.
By applying Eq.~\eqref{eqn:H_swap} and then applying Eq.~\eqref{eqn:dH_initial} with primed and unprimed variables switched, we have that
\begin{equation} \label{eqn:dH_final}
  \begin{split}
    \frac{\uD [(\tau' - \tau) \pbg H^{a'}{}_a]}{\ud \tau'} &= g_{ab} g^{a'b'} \frac{\uD [(\tau' - \tau) \pbg H^b{}_{b'}]}{\ud \tau'} \\
    &= g_{ab} g^{a'b'} \pbg K^b{}_{b'},
  \end{split}
\end{equation}
which is equivalent to Eq.~\eqref{eqn:prime_dH}.
Note that a similar type of analysis starting from Eq.~\eqref{eqn:K_swap} does not work, since the derivative on the right-hand side is with respect to $\tau$, and not $\tau'$.
However, by combining Eqs.~\eqref{eqn:KH_wronskian}, \eqref{eqn:H_swap}, and~\eqref{eqn:dH_final} and rearranging, it follows that
\begin{equation}
  (\tau' - \tau) \pbg H^a{}_{b'} \frac{\uD \pbg K^{b'}{}_b}{\ud \tau'} = \pbg K^a{}_{b'} \pbg K^{b'}{}_b - \delta^a{}_b,
\end{equation}
which is equivalent to Eq.~\eqref{eqn:prime_dK}.

\subsection{The (non-)geodesic deviation equation}

\subsubsection{Derivation}

We now turn to deriving a generalized geodesic deviation equation which allows for arbitrarily-accelerated worldlines (although we assume that the acceleration is perturbatively small).
This proof is based upon the discussion in~\cite{Vines2014a}, and generalizes the discussion in Paper~I to arbitrary order in separation and relative velocity.

In order to determine the deviation vector between two worldlines $\gamma$ and $\bar \gamma$ as a function of time, we first need to have a definition of this deviation vector.
The definition we use is given by the separation vector between two points, $x$ and $\bar x$, which lie on $\gamma$ and $\bar \gamma$, respectively.
In principle, we have a sort of ``gauge'' freedom to pick any rule for selecting the pairs of points $x$ and $\bar x$.
Following~\cite{Vines2014a}, we call this rule a \emph{correspondence}, and the choice that we make for much of this paper is to use the \emph{isochronous correspondence}: given an initial choice for $x$ and $\bar x$, we set the proper time $\tau$ for $\gamma$ and $\bar \gamma$ such that $x = \gamma(\tau)$ and $\bar x = \bar \gamma(\tau)$.
The pairs of points that we select along $\gamma$ and $\bar \gamma$ at later times are then those with equal values of proper time.
As such, we have that the separation vector is given, at all values of proper time $\tau$, by
\begin{equation}
  \xi^a \equiv -\sigma^a [\bar \gamma(\tau)].
\end{equation}

Since we ultimately want a differential equation for $\xi^a$, we take a derivative of $\xi^a$ with respect to $\tau$.
For any bitensor at $x$ and $\bar x$, such a derivative given by~\cite{Aleksandrov1979}
\begin{equation}
  \frac{\uD \Omega_{\ms A \bar{\ms B}}}{\ud \tau} = (\dot \gamma^c \nabla_c + \dot{\bar \gamma}^{\bar c} \nabla_{\bar c}) \Omega_{\ms A \bar{\ms B}}.
\end{equation}
For the case of $\xi^a$, this is
\begin{equation}
  \dot \xi^a = -\dot \gamma^b \sigma^a{}_b (\bar x) - \dot{\bar \gamma}^{\bar b} \sigma^a{}_{\bar b}.
\end{equation}
Using Eq.~\eqref{eqn:jacobi_def}, this can be written as
\begin{equation} \label{eqn:dot_gammabar}
  \dot{\bar \gamma}^{\bar a} = H^{\bar a}{}_a \dot \xi^a + K^{\bar a}{}_a \dot \gamma^a.
\end{equation}
While this expression is not useful as a differential equation for $\xi^a$ (it is in terms of $\dot{\bar \gamma}^{\bar a}$, which is unknown), it does mean that we can write
\begin{equation} \label{eqn:total_dtau}
  \frac{\uD \Omega_{\ms A \bar{\ms B}}}{\ud \tau} = (\dot \gamma^c \nabla_{c \bar *} + \dot \xi^c \nabla_{\bar * c}) \Omega_{\ms A \bar{\ms B}},
\end{equation}
where the \emph{horizontal} and \emph{vertical} derivatives $\nabla_{a \bar *}$ and $\nabla_{\bar * a}$ are defined by
\begin{equation}
  \nabla_{a \bar *} \equiv \nabla_a + K^{\bar a}{}_a \nabla_{\bar a}, \quad \nabla_{\bar * a} \equiv H^{\bar a}{}_a \nabla_{\bar a}
\end{equation}
(for a more geometric definition, see~\cite{Vines2014a, Dixon1979}).
Another application of these derivatives is that, since
\begin{equation}
  \nabla_{*'a} \sigma^b (x') = H^{a'}{}_a \sigma^b{}_{a'} = -\delta^b{}_a,
\end{equation}
we find that we can write Eq.~\eqref{eqn:coincident_expansion} in an alternative, more easily proven manner:
\begin{equation}
  \begin{split}
    \Omega_{\ms A} (x') = \sum_{n = 0}^\infty \frac{(-1)^n}{n!} &\sigma^{b_1} (x') \cdots \sigma^{b_n} (x') \\
    \times &[\nabla_{*'b_1} \cdots \nabla_{*'b_n} \Omega_{\ms A} (x')]_{x' \to x}.
  \end{split}
\end{equation}
The coincidence limit in this equation is known as the $n$th \emph{tensor extension} of $\Omega_{\ms A} (x')$, and is given by $n$ partial derivatives of $\Omega_{\ms A} (x')$ in a normal coordinate system around $x$~\cite{Dixon1979}.
Similarly, we have that
\begin{equation}
  \nabla_{a*'} \sigma^b (x') = \sigma^b{}_a (x') + K^{a'}{}_a \sigma^b{}_{a'} = 0.
\end{equation}

Applying Eq.~\eqref{eqn:total_dtau} once again to Eq.~\eqref{eqn:dot_gammabar}, we find that
\begin{equation} \label{eqn:ddot_gammabar}
  \begin{split}
    \ddot{\bar \gamma}^{\bar a} &= H^{\bar a}{}_a \ddot \xi^a + K^{\bar a}{}_a \ddot \gamma^a \\
    &\hspace{1.1em}+ I^{\bar a}{}_{ab} \dot \xi^a \dot \xi^b + 2 J^{\bar a}{}_{ab} \dot \xi^a \dot \gamma^b + L^{\bar a}{}_{ab} \dot \gamma^a \dot \gamma^b,
  \end{split}
\end{equation}
where
\begin{subequations}
  \begin{align}
    I^{\bar a}{}_{ab} &\equiv \nabla_{\bar * b} H^{\bar a}{}_a, \\
    J^{\bar a}{}_{ab} &\equiv \frac{1}{2} (\nabla_{\bar * a} K^{\bar a}{}_b + \nabla_{b \bar *} H^{\bar a}{}_{a}), \\
    L^{\bar a}{}_{ab} &\equiv \nabla_{b \bar *} K^{\bar a}{}_a.
  \end{align}
\end{subequations}
Given the accelerations of the two curves, Eq.~\eqref{eqn:ddot_gammabar} is in a desirable form; solving for $\ddot \xi^a$, we find that
\begin{equation}
  \begin{split}
    \ddot \xi^a &= -[\sigma^a{}_b (\bar x) \ddot \gamma^b + \sigma^a{}_{\bar b} \ddot{\bar \gamma}^{\bar b}] \\
    &\hspace{1.1em}+ \mc I^a{}_{bc} (\bar x) \dot \xi^b \dot \xi^c + 2 \mc J^a{}_{bc} (\bar x) \dot \xi^b \dot \gamma^c + \mc L^a{}_{bc} (\bar x) \dot \gamma^b \dot \gamma^c,
  \end{split}
\end{equation}
where
\begin{subequations}
  \begin{align}
    \mc I^a{}_{bc} (\bar x) &\equiv \sigma^a{}_{\bar a} I^{\bar a}{}_{bc}, \\
    \mc J^a{}_{bc} (\bar x) &\equiv \sigma^a{}_{\bar a} J^{\bar a}{}_{bc}, \\
    \mc L^a{}_{bc} (\bar x) &\equiv \sigma^a{}_{\bar a} L^{\bar a}{}_{bc}.
  \end{align}
\end{subequations}
As can be read off from Eqs.~(5.8) of~\cite{Vines2014a}, we have that
\begin{subequations}
  \begin{align}
    H^{\bar a}{}_a &= g^{\bar a}{}_a + O(\bs \xi)^2, \\
    K^{\bar a}{}_a &= g^{\bar a}{}_a + O(\bs \xi)^2, \\
    \mc I^a{}_{bc} (\bar x) &= O(\bs \xi), \\
    \mc J^a{}_{bc} (\bar x) &= O(\bs \xi), \\
    \mc L^a{}_{bc} (\bar x) &= -R^a{}_{bdc} \xi^d + O(\bs \xi^2),
  \end{align}
\end{subequations}
the first two of which imply that
\begin{subequations}
  \begin{align}
    \sigma^a{}_b (\bar x) &= \delta^a{}_b + O(\bs \xi^2), \\
    \sigma^a{}_{\bar b} &= -g^a{}_{\bar b} + O(\bs \xi^2).
  \end{align}
\end{subequations}
As such, we can write the general, non-geodesic deviation equation as
\begin{equation} \label{eqn:non_geodesic}
  \ddot \xi^a = -R^a{}_{\dot \gamma b \dot \gamma} \xi^b + a^a + O(\bs \xi, \dot{\bs \xi})^2,
\end{equation}
where
\begin{equation}
  \begin{split}
    a^a &\equiv -[\sigma^a{}_b (\bar x) \ddot \gamma^b + \sigma^a{}_{\bar b} \ddot{\bar \gamma}^{\bar b}] \\
    &= [g^a{}_{\bar a} + O(\bs{\xi}^2)] \ddot{\bar \gamma}^{\bar a} - [\delta^a{}_b + O(\bs{\xi}^2)] \ddot \gamma^b
  \end{split}
\end{equation}
is a notion of the relative acceleration of the two worldlines.

\subsubsection{Solution}

We now consider the solution to Eq.~\eqref{eqn:non_geodesic}.
First, we write this equation in the form
\begin{equation} \label{eqn:jacobi_inhomogeneous}
  \ddot \xi^a = -R^a{}_{\dot \gamma b \dot \gamma} \xi^b + S^a,
\end{equation}
where $S^a$ is some ``source'' term.
When one has $a^a \neq 0$, but neglects higher-order terms in separation and relative velocity, this is the equation one must solve directly, for $S^a = a^a$.
When one is instead considering higher-order corrections, this equation is obtained when solving Eq.~\eqref{eqn:non_geodesic} order-by-order: the source becomes a function of the lower-order solution.

The general solution is computed as follows~\cite{GrantThesis}: first, we note that this equation can be written in terms of the vector $X^A$ and the connection $\pbg \hat \nabla_a$ defined in Sec.~\ref{sec:jacobi} as
\begin{equation}
  \dot \gamma^b \pbg \hat \nabla_b X^A = S^A,
\end{equation}
where
\begin{equation}
  S^A \equiv \begin{pmatrix}
    0 \\
    S^a
  \end{pmatrix}.
\end{equation}
Now, note that the Leibniz rule and Eq.~\eqref{eqn:jacobi_init_deriv} imply that
\begin{equation}
  \frac{\ud}{\ud \tau'} (\pbg J^A{}_{A'} X^{A'}) = \pbg J^A{}_{A'} S^{A'},
\end{equation}
and so, integrating the left- and right-hand sides of these equations, we find that
\begin{equation}
  \pbg J^A{}_{A'} X^{A'} - X^A = \int_\tau^{\tau'} \ud \tau'' \pbg J^A{}_{A''} S^{A''},
\end{equation}
or
\begin{equation}
  X^{A'} = \pbg J^{A'}{}_A X^A + \int_\tau^{\tau'} \ud \tau'' \pbg J^{A'}{}_{A''} S^{A''},
\end{equation}
where we have used Eq.~\eqref{eqn:jacobi_composition}.
From this equation we now extract the first row, which, using Eq.~\eqref{eqn:jacobi_matrix}, yields
\begin{equation} \label{eqn:jacobi_inhomogeneous_soln}
  \begin{split}
    \xi^{a'} &= \pbg K^{a'}{}_a \xi^a + (\tau' - \tau) \pbg H^{a'}{}_a \dot \xi^a \\
    &\hspace{1.1em}+ \int_\tau^{\tau'} \ud \tau'' (\tau' - \tau'') \pbg H^{a'}{}_{a''} S^{a''}.
  \end{split}
\end{equation}
This equation appeared as Eq.~(4.10) of Paper~I, where it was proven by confirming that Eq.~\eqref{eqn:jacobi_inhomogeneous} was satisfied, instead of by introducing $\pbg J^{A'}{}_A$.

\section{Computation of persistent observables} \label{sec:persistent_observables}

We now discuss the computation of persistent observables, in terms of the various bitensors in the previous section (the parallel and Jacobi propagators).
Given such expressions, these observables can then be straightforwardly (if tediously) computed in any spacetimes where these bitensors are known.
While this is not always the most efficient method (for example, when the geodesic equation has known solutions, such as in the exact plane wave spacetimes considered in Paper~II), it does allow for more insight into results that apply to more general spacetimes.

\subsection{Curve deviation} \label{sec:curve_dev}

The first observable which we consider in this paper is the ``curve deviation'', which was introduced in Paper~I and studied at leading order in asymptotically flat spacetimes in Paper~III.
This is an observable that a pair of arbitrarily accelerating observers, following general worldlines $\gamma$ and $\bar \gamma$, can in principle measure, and arises as the most natural generalization of the displacement memory.

This pair of observers carries out the following procedure (see Paper~I): first, the observers establish their separation $\xi^a$ and relative velocity $\dot \xi^a$ at some initial proper time $\tau$.
At a later proper time $\tau'$, they then measure their separation $\xi^{a'}$; note that they make this measurement using the isochronous correspondence, associating points on their two worldlines which have the same value of proper time.
This would be a somewhat difficult procedure to do in a realistic experiment, but for simplicity we assume that this can be done by some set of ``ideal'' observers.
Moreover, the two observers should have tracked their accelerations as functions of time from $\tau$ until $\tau'$, which can easily be done with local accelerometers.
Once they have communicated this data to one another, they can then compute a predicted separation $\xi_{\rm flat}^{a'}$, based on the assumption that they had been traveling in a region of spacetime which is flat.
This predicted separation obeys the equation
\begin{equation}
  \ddot \xi_{\rm flat}^{a'} = a^{a'},
\end{equation}
which can be solved as a function of time to yield
\begin{equation}
  \begin{split}
    \xi_{\rm flat}^{a'} &= \pbg g^{a'}{}_a [\xi^a + (\tau' - \tau) \dot \xi^a] \\
    &\hspace{1.1em}+ \int_\tau^{\tau'} \ud \tau'' (\tau' - \tau'') \pbg g^{a'}{}_{a''} a^{a''}.
  \end{split}
\end{equation}
At this point, the observers now subtract this quantity from the true value of the separation $\xi^{a'}$, to obtain the curve deviation observable $\Delta \xi^{a'}$:
\begin{equation}
  \Delta \xi^{a'} \equiv \xi^{a'} - \xi_{\rm flat}^{a'}.
\end{equation}
This is non-zero only when the spacetime has had non-zero curvature at (not necessarily all) points between $\tau$ and $\tau'$, and so is a persistent observable in the context of the flat-to-flat transitions of Paper~I.
As was shown in Paper~III, it is also a persistent observable in asymptotically flat spacetimes, at $O(1/r)$.

We now discuss particular pieces of the curve deviation.
Suppose, for example, that $\gamma$ and $\bar \gamma$ are both geodesic, and that the initial relative velocity vanishes: this is the case in question when measuring the displacement memory.
To leading order, Eq.~\eqref{eqn:jacobi_inhomogeneous_soln} implies that
\begin{equation}
  \xi^{a'} = \pbg K^{a'}{}_a \xi^a + O(\bs \xi)^2,
\end{equation}
and so
\begin{equation}
  \Delta \xi^{a'} = \pbg \Delta K^{a'}{}_a \xi^a + O(\bs \xi)^2,
\end{equation}
where
\begin{equation}
  \pbg \Delta K^{a'}{}_a \equiv \pbg K^{a'}{}_a - \pbg g^{a'}{}_a.
\end{equation}
The quantity $\pbg \Delta K^{a'}{}_a$ is therefore the relevant quantity to compute when considering the displacement memory effect.

Consider now the addition of an initial relative velocity: in this case, Eq.~\eqref{eqn:jacobi_inhomogeneous_soln} instead shows that
\begin{equation}
  \Delta \xi^{a'} = \pbg \Delta K^{a'}{}_a \xi^a + (\tau' - \tau) \pbg \Delta H^{a'}{}_a \dot \xi^a + O(\bs \xi, \dot{\bs \xi})^2,
\end{equation}
where (similarly) we have
\begin{equation}
  \pbg \Delta H^{a'}{}_a \equiv \pbg H^{a'}{}_a - \pbg g^{a'}{}_a.
\end{equation}
This observable was previously called the ``subleading displacement memory''~\cite{Nichols2018, Flanagan2019a}, by the following logic: near null infinity, this effect can related to the ``spin''~\cite{Pasterski2015a, Nichols2017} and ``center-of-mass''~\cite{Nichols2018} memory effects.
These memory effects are ``subleading'', in the following three ways:
\begin{itemize}

\item in a post-Newtonian expansion~\cite{Nichols2017, Nichols2018, Siddhant2024} and in numerical relativity~\cite{Mitman2020, Grant2023}, they can be shown to be smaller in magnitude than the displacement memory effect for compact binary inspirals and mergers;

\item they can be related to ``conserved quantities'' (or ``charges'') that are constructed from pieces of the metric in Bondi coordinates at higher orders in $1/r$~\cite{Pasterski2015a, Nichols2018, Grant2021b}; and

\item these memory effects are related to ``subleading soft graviton theorems''~\cite{Pasterski2015a}.

\end{itemize}
However, using the term ``subleading'' when describing this observable in more general spacetimes is quite \emph{mis}leading: in such contexts, there is no reason why this effect needs to be smaller than the usually-considered displacement memory effect.
As such, we propose in this paper to rename it to ``drift memory'', as that more physically describes how it arises: it is a correction to the usual drifting apart of two observers with initial velocity relative to one another.

Before leaving behind the leading-order, geodesic case, we point out that there are two additional observables that can be computed from the curve deviation.
The first is the \emph{velocity} memory~\cite{Grishchuk1989}, which describes the dependence of the final relative velocity on the initial separation.
Note that, since $\uD \pbg \Delta K^{a'}{}_a/\ud \tau' = \uD \pbg K^{a'}{}_a/\ud \tau'$, one can recover the velocity memory either from directly looking at the final value of $\dot \xi^{a'}$ or from the first derivative of the curve deviation observable, $\uD \Delta \xi^{a'}/\ud \tau'$.
As the results of, for example, Paper~III imply (and as we will show explicitly below), the velocity memory vanishes in asymptotically flat spacetimes, when computed between two non-radiative regions.

Another velocity-related observable that has also been considered in the literature is the final relative velocity as a function of initial velocity~\cite{Seraj2021a, Divakarla2021}.
Adopting the terminology of~\cite{Seraj2021a} (which may be due to the terminology appearing in the literature on electromagnetic memory~\cite{Bieri2013}\footnote{Kick memory was also observed in~\cite{Ferko2021} in the context of purely gravitational effects in spacetimes with compact extra dimensions, where the relationship with electromagnetic and color memory is more apparent.}), we call this the \emph{kick} memory.
This observable is given by $\uD [(\tau' - \tau) \pbg \Delta H^{a'}{}_a]/\ud \tau'$, and it is known that this effect is non-zero in asymptotically flat spacetimes~\cite{Seraj2021a, Divakarla2021}.\footnote{This provides another reason why using the term ``subleading displacement memory'' to describe the drift memory is problematic: by analogy, it motivates referring to the kick memory as the ``subleading velocity memory'', when in fact it is \emph{super}leading in asymptotically flat spacetimes!}
As we will show in Sec.~\ref{sec:redundant}, the kick memory is somewhat redundant, and in perturbative contexts it can be entirely determined from the displacement, drift, and velocity memories.
This generalizes results discussed in Paper~II.

We consider now the effect of adding acceleration terms.
Here, we are solving Eq.~\eqref{eqn:jacobi_inhomogeneous} with $S^a = a^a$, and so it follows that
\begin{equation} \label{eqn:curve_deviation_a}
  \begin{split}
    \Delta \xi^{a'} &= \pbg \Delta K^{a'}{}_a \xi^a + \pbg \Delta H^{a'}{}_a \dot \xi^a \\
    &\hspace{1.1em}+ \int_\tau^{\tau'} \ud \tau''\; (\tau' - \tau) \pbg \Delta H^{a'}{}_{a''} a^{a''} + O(\bs \xi, \dot{\bs \xi}, \bs a)^2.
  \end{split}
\end{equation}
Moreover, as it can be done without loss of generality (see Appendix~B of Paper~II), we focus exclusively on the case where only $\bar \gamma$ is accelerating, and so
\begin{equation}
  a^a = [g^a{}_{\bar a} + O(\bs \xi^2)] \ddot{\bar \gamma}^{\bar a}.
\end{equation}
As was shown in Appendix~C of Paper~III, when $a^{a''}$ can be Taylor-expanded in powers of $\tau'' - \tau$ (for any $\tau''$ between $\tau$ and $\tau'$), one has the following expansion in initial derivatives of $\dot{\bar \gamma}^{\bar a}$:
\begin{equation}
  a^{a''} = \pbg g^{a''}{}_a g^a{}_{\bar a} \sum_{n = 0}^\infty \frac{(\tau'' - \tau)^n}{n!} \frac{\uD^n \ddot{\bar \gamma}^{\bar a}}{\ud \tau^n} + O(\bs \xi, \dot{\bs \xi}, \ddot{\bar{\bs \gamma}})^2.
\end{equation}
As such, Eq.~\eqref{eqn:curve_deviation_a} can be written as
\begin{equation} \label{eqn:curve_deviation_init}
  \begin{split}
    \Delta \xi^{a'} &= \pbg \Delta K^{a'}{}_a \xi^a + \pbg \Delta H^{a'}{}_a \dot \xi^a \\
    &\hspace{1.1em}+ \sum_{n = 0}^\infty \pbg \Delta \ord{n} \alpha^{a'}{}_a g^a{}_{\bar a} \frac{\uD^n \ddot{\bar \gamma}^{\bar a}}{\ud \tau^n} + O(\bs \xi, \dot{\bs \xi}, \ddot{\bar{\bs \gamma}})^2,
  \end{split}
\end{equation}
where
\begin{equation} \label{eqn:Deltaalpha}
  \pbg \Delta \ord{n} \alpha^{a'}{}_a \equiv \frac{1}{n!} \int_\tau^{\tau'} \ud \tau'' (\tau'' - \tau)^n (\tau' - \tau'') \pbg \Delta H^{a'}{}_{a''} \pbg g^{a''}{}_a. \vspace{1em}
\end{equation}
In Paper~III, $\pbg \Delta K^{a'}{}_a$, $\pbg \Delta H^{a'}{}_a$, and $\pbg \Delta \ord{n} \alpha^{a'}{}_a$ were all computed in asymptotically flat spacetimes in terms of (temporal) moments of the Bondi news.
We present a generalization of this result below in Sec.~\ref{sec:moments}.
We refer to $\pbg \Delta \ord{n} \alpha^{a'}{}_a$, interchangeably with the moments of the news from which it can be computed, as being related to \emph{higher memories}.
We propose the name ``ballistic memory'' for the zeroth higher memory, as $\pbg \Delta \ord{0} \alpha^{a'}{}_a$ will be relevant for the first time if the observers experience constant relative acceleration, as occurs in projectile motion in classical mechanics.

\subsection{Nonlinear relationships between observables} \label{sec:nonlinear}

\subsubsection{Redundancy of the kick memory} \label{sec:redundant}

We first show that the Wro\'nskian in Eq.~\eqref{eqn:KH_wronskian} can be used to relate different pieces of the curve deviation and its derivative.
That is, we show that the displacement, velocity, drift, and kick memories are not all independent.
We start by expanding
\begin{widetext}
\begin{equation}
  g_{a'b'} \pbg K^{b'}{}_b \frac{\uD [(\tau' - \tau) \pbg H^{a'}{}_a]}{\ud \tau'} = g_{ab} + g_{a'b'} \left\{\pbg \Delta K^{b'}{}_b \pbg g^{a'}{}_a + \pbg K^{b'}{}_b \frac{\uD [(\tau' - \tau) \pbg \Delta H^{a'}{}_a]}{\ud \tau'}\right\},
\end{equation}
and upon inserting this equation into Eq.~\eqref{eqn:KH_wronskian}, we find
\begin{equation} \label{eqn:Delta_Wronskian}
  g_{a'b'} \pbg K^{b'}{}_b \frac{\uD [(\tau' - \tau) \pbg \Delta H^{a'}{}_a]}{\ud \tau'} = g_{a'b'} \left[(\tau' - \tau) \pbg H^{a'}{}_a \frac{\uD \pbg K^{b'}{}_b}{\ud \tau'} - \pbg \Delta K^{b'}{}_b \pbg g^{a'}{}_a\right].
\end{equation}
As such, if $\pbg K^{a'}{}_a$ is invertible, then we can write the kick memory in terms of the other three memory effects:
\begin{equation} \label{eqn:kick_redundant}
  \frac{\uD [(\tau' - \tau) \pbg \Delta H^{a'}{}_a]}{\ud \tau'} = (\pbg K^{-1})^b{}_{b'} g^{a'b'} g_{c'd'} \left[(\tau' - \tau) (\pbg g^{c'}{}_a + \pbg \Delta H^{c'}{}_a) \frac{\uD \pbg \Delta K^{d'}{}_b}{\ud \tau'} - \pbg g^{c'}{}_a \pbg \Delta K^{d'}{}_b\right].
\end{equation}
\end{widetext}

There are many situations where $\pbg K^{a'}{}_a$ is not invertible.
For example, when $x$ and $x'$ are conjugate points along $\gamma$, then for any $\tau'' \in (\tau, \tau')$, $\pbg K^{a''}{}_a$ and $\pbg K^{a'}{}_{a''}$ cannot be invertible.\footnote{Note that the definition of $x$ and $x'$ being conjugate points (see, for example, Section~5.3 of~\cite{DoCarmo1992}) is that there exists a non-zero Jacobi field that vanishes at $x$ and $x'$, or in other words that there exists a vector field $A^a$ such that $(\tau' - \tau) \pbg H^{a'}{}_a A^a = 0$.
  It does not necessarily imply that there are multiple geodesics connecting $x$ and $x'$, unless one is in a situation where the geodesic deviation equation has no higher-order corrections.}
However, a sufficient condition that $\pbg K^{a'}{}_a$ be invertible is given by the following: define
\begin{equation}
  \pbg \Delta \mc K^a{}_b (\tau') \equiv \pbg g^a{}_{a'} \pbg \Delta K^{a'}{}_b,
\end{equation}
and note that
\begin{equation}
  \pbg \mc K^a{}_b (\tau') \equiv \pbg g^a{}_{a'} \pbg K^{a'}{}_b = \delta^a{}_b + \pbg \Delta \mc K^a{}_b (\tau').
\end{equation}
As such, if $\pbg \Delta \mc K^a{}_b (\tau')$ is ``small'' in the sense that
\begin{equation} \label{eqn:DeltaK_small}
  \lim_{n \to \infty} (\pbg \Delta \mc K^n)^a{}_b (\tau') = 0,
\end{equation}
then
\begin{equation}
  \begin{split}
    (\pbg K^{-1})^a{}_{a'} &= (\pbg \mc K^{-1})^a{}_b \pbg g^b{}_{a'} \\
    &= \pbg g^b{}_{a'} \sum_{n = 0}^\infty (-1)^n (\pbg \Delta \mc K^n)^a{}_b (\tau').
  \end{split}
\end{equation}
Equation~\eqref{eqn:DeltaK_small}, and therefore Eq.~\eqref{eqn:kick_redundant}, will hold in a perturbative context, such as in an expansion in $1/r$.

\subsubsection{Bitensorial moments} \label{sec:moments}

In the previous section, Eq.~\eqref{eqn:Delta_Wronskian} provided a pointwise relationship between the observables arising at linear order for geodesics, implying the existence of essentially only three point-wise independent observables.
However, if one is allowed to know the values of these observables at all values of their arguments, then their derivatives are not independent either, implying that there are only two independent observables (say, displacement and drift).
Moreover, Eq.~\eqref{eqn:dH_initial} implies that
\begin{equation}
  (\tau' - \tau) \pbg H^{a'}{}_a = \int_\tau^{\tau'} \ud \tau''\; \pbg K^{a'}{}_{a''} \pbg g^{a''}{}_a,
\end{equation}
and so
\begin{equation} \label{eqn:DeltaH_DeltaK}
  (\tau' - \tau) \pbg \Delta H^{a'}{}_a = \int_\tau^{\tau'} \ud \tau''\; \pbg \Delta K^{a'}{}_{a''} \pbg g^{a''}{}_a.
\end{equation}
As such, in this same sense there is really only \emph{one} independent observable.

Using this type of argument, we can show, in fact, that all of the linear curve deviation observables can be written in terms of integrals of a single bitensor.
A similar result, not in terms of bitensors, was shown in Paper~III: under appropriate circumstances, all of these observables could be written in terms of the (temporal) moments of the tensor associated with radiation in asymptotically flat spacetimes: the Bondi news~\cite{Bondi1962, Sachs1962b, Geroch1977}.
The existence of a nonlinear generalization is not immediately obvious, and it implies that one could in principle easily extend the results of Paper~III to $O(1/r^2)$.

We start this analysis by noting that, by multiplying Eq.~\eqref{eqn:dK_initial} by $\pbg g^a{}_{b'}$,
\begin{equation}
  \frac{\ud}{\ud \tau} (\pbg K^{a'}{}_a \pbg g^a{}_{b'}) = (\tau' - \tau) \pbg H^{a'}{}_b R^b{}_{\dot \gamma a \dot \gamma} \pbg g^a{}_{b'};
\end{equation}
integrating this equation and inverting the parallel propagator yields
\begin{equation} \label{eqn:K_N}
  \pbg K^{a'}{}_a = \pbg g^{a'}{}_a + \int_\tau^{\tau'} \ud \tau''\; \pbg E^{a'}{}_a (\tau''),
\end{equation}
where
\begin{equation} \label{eqn:N}
  \pbg E^{a'}{}_a (\tau'') \equiv -(\tau' - \tau'') \pbg H^{a'}{}_{a''} R^{a''}{}_{\dot \gamma'' b'' \dot \gamma''} \pbg g^{b''}{}_a.
\end{equation}
This object is a bitensorial modification of the electric part of the Riemann tensor, hence the use of the letter `$E$'.
This immediately implies that
\begin{equation} \label{eqn:zeroth_moment}
  \pbg \Delta K^{a'}{}_a = \int_\tau^{\tau'} \ud \tau''\; \pbg E^{a'}{}_a (\tau''),
\end{equation}
providing a direct generalization of Eq.~(3.10) of Paper~III.
Next, consider the following identity, for some $f^{a'}{}_a (\tau'')$:
\begin{widetext}
\begin{equation}
  \frac{\ud}{\ud \tau''} \left[\frac{(\tau'' - \tau)^{n + 1}}{n + 1} \int_{\tau''}^{\tau'} \ud \tau'''\; f^{a'}{}_a (\tau'')\right] = (\tau'' - \tau)^n \int_{\tau''}^{\tau'} \ud \tau''' f^{a'}{}_a (\tau'') - \frac{(\tau'' - \tau)^{n + 1}}{n + 1} f^{a'}{}_a (\tau''),
\end{equation}
and so, since the term in brackets on the left-hand side vanishes when $\tau'' = \tau$ or $\tau'$, we find that
\begin{equation} \label{eqn:cauchy_identity}
  \int_\tau^{\tau'} \ud \tau''\; (\tau'' - \tau)^n \int_{\tau''}^{\tau'} \ud \tau''' f^{a'}{}_a (\tau'') = \frac{1}{n + 1} \int_\tau^{\tau'} \ud \tau''\; (\tau'' - \tau)^{n + 1} f^{a'}{}_a (\tau'').
\end{equation}
This all relies upon the fact that we are integrating $f^{a'}{}_a (\tau'')$ over the points over which it is a scalar, and so the usual intuition from single-variable calculus applies.
We can now apply this equation in the case $f^{a'}{}_a (\tau'') = \pbg E^{a'}{}_a (\tau'')$.
By using Eq.~\eqref{eqn:DeltaH_DeltaK}, together with this equation for $n = 0$, we find that
\begin{equation} \label{eqn:first_moment}
  (\tau' - \tau) \pbg \Delta H^{a'}{}_a = \int_\tau^{\tau'} \ud \tau''\; (\tau'' - \tau) \pbg E^{a'}{}_a (\tau''),
\end{equation}
a generalization of Eq.~(3.13) of Paper~III.

Finally, using Eq.~\eqref{eqn:DeltaH_DeltaK}, Eq.~\eqref{eqn:Deltaalpha} becomes
\begin{equation} \label{eqn:Deltaalpha_DeltaK}
  \pbg \Delta \ord{n} \alpha^{a'}{}_a = \frac{1}{n!} \int_\tau^{\tau'} \ud \tau''\; (\tau'' - \tau)^n \int_{\tau''}^{\tau'} \ud \tau''' \pbg \Delta K^{a'}{}_{a'''} \pbg g^{a'''}{}_a.
\end{equation}
Applying Eq.~\eqref{eqn:cauchy_identity} to Eq.~\eqref{eqn:Deltaalpha_DeltaK}, where $f^{a'}{}_a (\tau'') = \pbg \Delta K^{a'}{}_{a''} \pbg g^{a''}{}_a$ [note that this is still a scalar at $\gamma(\tau'')$!], we find
\begin{equation}
  \pbg \Delta \ord{n} \alpha^{a'}{}_a = \frac{1}{(n + 1)!} \int_\tau^{\tau'} \ud \tau''\; (\tau'' - \tau)^{n + 1} \int_{\tau''}^{\tau'} \ud \tau'''\; \pbg E^{a'}{}_a (\tau'''),
\end{equation}
\end{widetext}
by applying Eq.~\eqref{eqn:zeroth_moment} and the fact that
\begin{equation}
  \pbg E^{a'}{}_{a''} (\tau''') \pbg g^{a''}{}_a = \pbg E^{a'}{}_a (\tau''').
\end{equation}
Therefore, applying Eq.~\eqref{eqn:cauchy_identity} once again yields
\begin{equation}
  \pbg \Delta \ord{n} \alpha^{a'}{}_a = \frac{1}{(n + 2)!} \int_\tau^{\tau'} \ud \tau''\; (\tau'' - \tau)^{n + 2} \pbg E^{a'}{}_a (\tau''),
\end{equation}
a direct generalization of Eq.~(3.14) of Paper~III.
As such, denoting by
\begin{equation}
  \pbg \ord{n}{\ms E}^{a'}{}_a \equiv \frac{1}{n!} \int_\tau^{\tau'} \ud \tau''\; (\tau'' - \tau)^n \pbg E^{a'}{}_a (\tau'')
\end{equation}
the $n$th (temporal) moment of the bitensor $\pbg E^{a'}{}_a (\tau'')$, we find that
\begin{subequations} \label{eqn:N_moments}
  \begin{align}
    \pbg \Delta K^{a'}{}_a &= \pbg \ord{0}{\ms E}^{a'}{}_a, \\
    (\tau' - \tau) \pbg \Delta H^{a'}{}_a &= \pbg \ord{1}{\ms E}^{a'}{}_a, \\
    \pbg \Delta \ord{n} \alpha^{a'}{}_a &= \pbg \ord{n + 2}{\ms E}^{a'}{}_a.
  \end{align}
\end{subequations}

Note that one can recover the velocity-related observables (such as the velocity and kick memories) by taking derivatives of these expressions.
This suggests that it is useful to consider a parallel set of moments, $\pbg \ord{n}{\widetilde{\ms E}}^{a'}{}_a$, defined by
\begin{equation}
  \begin{split}
    \pbg \ord{n}{\widetilde{\ms E}}^{a'}{}_a &\equiv \frac{\uD}{\ud \tau'} \pbg \ord{n}{\ms E}^{a'}{}_a \\
    &= \frac{1}{n!} \int_\tau^{\tau'} \ud \tau''\; (\tau'' - \tau)^n \pbg \widetilde E^{a'}{}_a (\tau''),
  \end{split}
\end{equation}
using the fact that $\pbg E^{a'}{}_a (\tau') = 0$, and where
\begin{equation} \label{eqn:N_tilde}
  \pbg \widetilde E^{a'}{}_a (\tau'') \equiv -g^{a'b'} \pbg K^{a''}{}_{b'} R_{a''\dot \gamma'' b'' \dot \gamma''} \pbg g^{b''}{}_a,
\end{equation}
which uses Eq.~\eqref{eqn:dH_final}.

We now show how one can easily derive the results of Paper~III by using the results of this section.
First, note that, in Bondi coordinates $\{u, r, \theta^i\}$, the Riemann tensor takes the following form:
\begin{equation} \label{eqn:asymp_R}
  R^i{}_{uju} = -\frac{1}{2r} \partial_u N^i{}_j + O(1/r^2),
\end{equation}
where $N_{ij}$ is the \emph{news tensor}, and angular coordinate indices (denoted by Latin letters from the middle of the alphabet) are raised and lowered using the metric on the round, unit two-sphere.
As was shown explicitly in Paper~III, one can have asymptotic observers whose four-velocity $\dot \gamma^a = (\partial_u)^a + O(1/r)$, where ``$O(1/r)$'' refers to subleading terms when this vector is expanded on an orthonormal basis.
As such, at leading order we are free to use $u$ as our affine parameter, and moreover, in Paper~III it was shown that $\frac{1}{r} (\partial_i)^a$ and $r (\ud \theta^i)_a$ are parallel-transported along this curve to leading order.
Therefore, using Eqs.~\eqref{eqn:N} and~\eqref{eqn:asymp_R}, it follows that at leading order the angular components of $\pbg E^{a'}{}_a (\tau'')$ are given by
\begin{equation}
  \pbg E^i{}_j (u', u; u'') = \frac{1}{2r} (u' - u'') \partial_{u''} N^i{}_j (u'').
\end{equation}
Next, writing
\begin{equation}
  u' - u'' = (u' - u) - (u'' - u),
\end{equation}
it follows that, for any function $f$ which vanishes at $u$ and $u'$,
\begin{widetext}
\begin{equation}
  \begin{split}
    \frac{1}{n!} \int_u^{u'} \ud u'' (u'' - u)^n (u' - u'') \dot f(u'') &= \frac{1}{n!} \left[(u' - u) \int_u^{u'} \ud u'' (u'' - u)^n \dot f(u'') - \int_u^{u'} \ud u'' (u'' - u)^{n + 1} \dot f(u'')\right] \\
    &= \frac{n + 1}{n!} \int_u^{u'} \ud u' (u'' - u)^n f(u'') - \begin{dcases}
      0 & n = 0 \\
      \frac{u' - u}{(n - 1)!} \int_u^{u'} \ud u'' (u'' - u)^{n - 1} f(u'') & n \neq 0
    \end{dcases},
  \end{split}
\end{equation}
where the second line follows by an ordinary integration by parts.
As such, we find that, in the case where $N_{ij}$ is assumed to vanish at $u$ and $u'$ (as was done in Paper~III),
\begin{equation} \label{eqn:N_asymp}
  \pbg \ord{n}{\ms E}^i{}_j (u', u) = \frac{1}{2r} \left[(n + 1) \ord{n}{\mc N}^i{}_j (u', u) - (u' - u) \begin{dcases}
    0 \vphantom{\overset{_{(n - 1)}}{\mc N}} & n = 0 \\
    \ord{n - 1}{\mc N}^i{}_j (u', u) & n > 0
  \end{dcases}\right] + O(1/r^2),
\end{equation}
\end{widetext}
where
\begin{equation}
  \ord{n}{\mc N}^i{}_j (u', u) \equiv \frac{1}{n!} \int_u^{u'} \ud u''\; (u'' - u)^n N^i{}_j (u'')
\end{equation}
is the $n$th moment of the news, as defined in Paper~III (see~\cite{Grant2023, Siddhant2024} for other definitions that are more useful in the contexts of those papers).
Combining Eqs.~\eqref{eqn:N_moments} and~\eqref{eqn:N_asymp} directly proves Eqs.~(3.12-14) of Paper~III, and this procedure provides a straightforward approach for computing the extensions of those equations to higher order in $1/r$.

We can also perform a similar computation for the derivatives of these moments.
First, using Eqs.~\eqref{eqn:N_tilde} and~\eqref{eqn:asymp_R}, we find that
\begin{equation}
  \pbg \widetilde E^i{}_j (u', u; u'') = \frac{1}{2r} \partial_u N^i{}_j (u'') + O(1/r^2).
\end{equation}
By using a similar integration-by-parts procedure as was used to derive Eq.~\eqref{eqn:N_asymp}, we recover
\begin{equation}
  \begin{split}
    \pbg \ord{n}{\widetilde{\ms E}}^i{}_j (u', u) &= -\frac{1}{2r} \begin{cases}
      0 & n = 0 \\
      \ord{n - 1}{\mc N}^i{}_j (u', u) & n > 0
    \end{cases} \\
    &\hspace{1.1em}+ O(1/r^2).
  \end{split}
\end{equation}
This equation shows that, at leading order in $1/r$, the velocity memory vanishes and the kick memory is (apart from a sign) the same as the displacement memory, as discussed above in Sec.~\ref{sec:curve_dev}.

\subsection{Proper time shift observable} \label{sec:proper_time}

We next turn to the proper time shift observable, which arises in a setup which is slightly different from the curve deviation.
Here, instead of the two observers associating points on their two worldlines by enforcing that they have equal values of proper time, they associate points such that the separation vector, which we denote by $\xi_\perp^a$, is orthogonal to $\gamma$, for all times $\tau$:
\begin{equation} \label{eqn:normal}
  \xi_\perp^a \dot \gamma_a = 0.
\end{equation}
This is the so-called \emph{normal} correspondence~\cite{Vines2014a}.

Note that Eq.~\eqref{eqn:jacobi_final_u} implies that, if $\xi^a \dot \gamma_a = 0$ initially, then, assuming that there is no acceleration or initial relative velocity,
\begin{equation}
  \xi^{a'} \dot \gamma_{a'} = O(\bs \xi)^2
\end{equation}
at all later times $\tau'$.
Moreover, when one considers the curve deviation observable, Eqs.~\eqref{eqn:jacobi_final_u} and~\eqref{eqn:curve_deviation_init} imply that
\begin{equation}
  \Delta \xi^{a'} \dot \gamma_{a'} = O(\ddot{\bs \gamma}, \ddot{\bar{\bs \gamma}}) + O(\bs \xi, \dot{\bs \xi})^2.
\end{equation}
As such, it follows that any interesting differences between $\xi_\perp^a$ and $\xi^a$ must arise at second order, or from the presence of acceleration terms.

Since $\xi_\perp^a$ is not defined using the isochronous correspondence, it no longer points between $x \equiv \gamma(\tau)$ and $\bar x \equiv \bar \gamma(\tau)$, but $x$ and $\hat{\bar x} \equiv \bar \gamma(\hat \tau)$, where $\hat \tau \equiv \tau + \Delta \tau$: the proper time shift observable is this quantity $\Delta \tau$.
To compute $\Delta \tau$, we first add to this list of points $\hat x \equiv \gamma(\tau + \Delta \tau)$, and consider the triangle composed of $x$, $\hat x$, and $\hat{\bar x}$.
Assuming for simplicity that $\gamma$ is a geodesic, and that $\Delta \tau$ is small enough that there is only one geodesic between $x$ and $\hat x$, then
\begin{equation}
  \Delta \tau \dot \gamma^{\hat a} = \sigma^{\hat a} (x).
\end{equation}
We can now determine $\xi_\perp^a$ by applying a procedure presented in~\cite{Vines2014b}: first, define
\begin{equation}
  w^{\hat a} \equiv g^{\hat a}{}_a \xi_\perp^a,
\end{equation}
and then expand in powers of $\xi^{\hat a}$ and $-\Delta \tau \dot \gamma^{\hat a}$.

Note that, for this entire procedure to work, we need to know that $\Delta \tau$ is itself small, which is not immediately clear.
However, this holds by the following argument: first, in the case where $\bar \gamma$ is a geodesic, note that when $\gamma$ and $\bar \gamma$ intersect at $x$ (which corresponds to $\xi^a = 0$), then $\dot{\bar \gamma}^a$ and $\xi_\perp^a$ must be colinear, as $\xi_\perp^a$ points from $x$ to $\hat{\bar x}$ along the unique geodesic between them, and $\bar \gamma$ is such a geodesic (note that, if there are multiple geodesics, $\xi_\perp^a$ is not defined, so the question is moot).
Since $\xi_\perp^a \dot \gamma_a = 0$, this implies that $\xi_\perp^a$ must vanish when $\xi^a$ vanishes.  However, this means that $\Delta \tau$ is zero as well, and so $\Delta \tau$ and $\xi_\perp^a$ are both $O(\bs \xi)$, and so are small.
When $\bar \gamma$ is \emph{not} a geodesic, $\Delta \tau$ is therefore $O(\bs \xi, \ddot{\bar{\bs \gamma}})$, which is also small.

Since we can assume that $\Delta \tau$ is small, we can apply Eq.~(37) of~\cite{Vines2014b}, obtaining
\begin{equation}
  w^{\hat a} = -\Delta \tau \dot \gamma^{\hat a} - \xi^{\hat a} + O(\Delta \tau, \bs \xi)^3.
\end{equation}
Applying Eq.~\eqref{eqn:curve_deviation_a}, \eqref{eqn:zeroth_moment}, and~\eqref{eqn:first_moment}, we also find that
\begin{equation}
  \xi^{\hat a} = g^{\hat a}{}_a \xi^a + \Delta \tau g^{\hat a}{}_a \dot \xi^a + O(\Delta \tau, \bs \xi, \dot{\bs \xi}, \ddot{\bar{\bs \gamma}})^3. \vspace{1em}
\end{equation}
Using Eq.~\eqref{eqn:normal} and the fact that $\dot \gamma^a$ is parallel-transported, we therefore find that
\begin{equation}
  \begin{split}
    0 &= \dot \gamma_{\hat a} w^{\hat a} \\
    &= \Delta \tau - \xi^a \dot \gamma_a - \Delta \tau \dot \xi^a \dot \gamma_a + O(\Delta \tau, \bs \xi, \dot{\bs \xi}, \ddot{\bar{\bs \gamma}})^3.
  \end{split}
\end{equation}
By solving this equation iteratively for $\Delta \tau$, our final result is
\begin{equation}
  \Delta \tau = \frac{\xi^a \dot \gamma_a}{1 - \dot \gamma_a \dot \xi^a} + O(\bs \xi, \dot{\bs \xi}, \ddot{\bar{\bs \gamma}})^3.
\end{equation}

Considering this equation at $\tau'$, we need both $\xi^{a'} \dot \gamma_{a'}$ and $\dot \xi^{a'} \dot \gamma_{a'}$ in order to compute $\Delta \tau'$.
As we are assuming that $\gamma$ is a geodesic, the latter is just the derivative of the former with respect to $\tau'$.
Using Eq.~\eqref{eqn:jacobi_inhomogeneous_soln}, we can give $\xi^{a'} \dot \gamma_{a'}$ in terms of the arbitrary source $S^a$ appearing on the right-hand side of Eq.~\eqref{eqn:jacobi_inhomogeneous}:
\begin{equation}
  \begin{split}
    \xi^{a'} \dot \gamma_{a'} &= [\xi^a + (\tau' - \tau) \dot \xi^a] \dot \gamma_a \\
    &\hspace{1.1em}+ \int_\tau^{\tau'} \ud \tau''\; (\tau' - \tau'') \dot \gamma_{a''} S^{a''},
  \end{split}
\end{equation}
from which it follows that
\begin{equation}
  \dot \xi^{a'} \dot \gamma_{a'} = \dot \xi^a \dot \gamma_a + \int_\tau^{\tau'} \ud \tau''\; \dot \gamma_{a''} S^{a''}.
\end{equation}

We now compute $\Delta \tau'$ in the concrete case considered in Paper~I, namely where $\xi^a \dot \gamma_a = 0$, and there is neither initial relative velocity nor acceleration.
As such, we recover
\begin{equation} \label{eqn:Deltatau_S}
  \Delta \tau' = \int_\tau^{\tau'} \ud \tau''\; (\tau' - \tau'') \dot \gamma_{a''} S^{a''} + O(\bs \xi, \dot{\bs \xi})^3.
\end{equation}
From Eq.~(4.5) of Paper~I, it follows that
\begin{equation}
  \begin{split}
    S^{a'} &= -\left(2 R^{a'}{}_{c'b' \dot \gamma'} \dot \xi^{c'} + \nabla_{(\dot \gamma'} R^{a'}{}_{c')b' \dot \gamma'} \xi^{c'}\right) \xi^{b'} \\
    &\hspace{1.1em}+ O(\bs \xi, \dot{\bs \xi})^2,
  \end{split}
\end{equation}
where, like in Eq.~\eqref{eqn:dotgamma_contract} we are using $\dot \gamma$ as an index to indicate contraction with the four-velocity.
As such, using Eq.~\eqref{eqn:jacobi_inhomogeneous_soln} we find that
\begin{widetext}
\begin{equation}
  \dot \gamma_{a'} S^{a'} = \frac{1}{2} \pbg K^{b'}{}_b \xi^a \xi^b \left[\frac{\uD}{\ud \tau'} \left(\pbg K^{a'}{}_a R_{a' \dot \gamma' b' \dot \gamma'}\right) + 3 R_{a' \dot \gamma' b' \dot \gamma'} \frac{\uD \pbg K^{a'}{}_a}{\ud \tau'}\right] + O(\bs \xi)^3.
\end{equation}
Using Eq.~\eqref{eqn:Deltatau_S} and an integration by parts, we therefore recover that
\begin{equation} \label{eqn:Deltatau_final}
  \Delta \tau' = \frac{1}{2} \left\{(\tau' - \tau) R_{a \dot \gamma b \dot \gamma} + \int_\tau^{\tau'} \ud \tau''\; R_{a'' \dot \gamma'' b'' \dot \gamma''} \left[\pbg K^{a''}{}_a + 3 (\tau' - \tau'') \frac{\uD \pbg K^{a''}{}_a}{\ud \tau'}\right] \pbg K^{b''}{}_b\right\} \xi^a \xi^b + O(\bs \xi)^3.
\end{equation}
\end{widetext}
The first term can be neglected if the initial Riemann tensor is set to zero (as was the case in Paper~I).
Moreover, when one expands perturbatively in the Riemann tensor, Eq.~\eqref{eqn:K_N} implies that
\begin{equation}
  \pbg K^{a'}{}_a = \pbg g^{a'}{}_a + O(\bs R).
\end{equation}
In such an expansion, the second term in the square brackets can therefore also be neglected, and so we recover Eq.~(2.6) of Paper~I, which was given without proof.
Neglecting the first term, Eq.~\eqref{eqn:Deltatau_final} is the generalization to the case where the curvature is not assumed to be small.

\section{Discussion} \label{sec:discussion}

In this paper, we have considered nonlinear effects arising in persistent observables, which are usually studied in asymptotically flat spacetimes, in an effort both to understand these observables more deeply and to yield results which will be applicable beyond the leading order in $1/r$.
The properties of the Jacobi propagators in general spacetimes were crucial in the derivation of these results, and so we reviewed them in detail.
Using these properties, along with general expressions for the solutions to the equation of (non-)geodesic deviation, we analyzed the following two observables:
\begin{itemize}

\item the curve deviation observable of Paper~I, which contains the usual (displacement) memory effect, along with the drift, ballistic, and higher memory effects; and

\item a proper time shift observable considered in~\cite{Strominger2014b} and Paper~I.

\end{itemize}
For the former, we first considered pointwise relationships between the Jacobi propagators, which showed that the displacement, drift, velocity, and kick memories were not all independent.
Next, we considered differential relationships between the Jacobi propagators, and showed that they allowed for a generalization of the results of Papers~I and~III, namely that the different pieces of the curve deviation can be written in terms of appropriately-defined moments.
Finally, we provided an explicit proof of the results of Paper~I for the proper time observable, and since we performed the calculation using bitensors, we also extended them beyond linear order in the curvature.

With the completion of the analysis of the curve deviation in terms of nonlinearly-defined moments, it would be interesting to determine whether a similar analysis can be performed for the remaining observables that were defined in Paper~I.
In particular, it would be useful to understand how to write the holonomy observable in terms of moments.
This observable was defined using the map which took vectors (sections of some higher-dimensional vector bundle, motivated by~\cite{Flanagan2016}) and parallel-transported them, with respect to some connection $\widetilde \nabla_a$, around a loop formed by two closely-separated worldlines and the unique geodesics between them.
Denoting this map by $\widetilde \Lambda^A{}_B$ (where, as in Sec.~\ref{sec:jacobi}, capital Latin indices denote indices on this arbitrary vector bundle), the holonomy observable was given by
\begin{equation}
  \begin{split}
    \widetilde \Omega^A{}_B &\equiv \widetilde \Lambda^A{}_B - \delta^A{}_B \\
    &= \int_\tau^{\tau'} \ud \tau'' \pbg \widetilde g^A{}_{A''} \widetilde R^{A''}{}_{B''c''d''} \xi^{c''} \dot \gamma^{d''} \pbg \widetilde g^{B''}{}_B,
  \end{split}
\end{equation}
where $\pbg \widetilde g^{A'}{}_A$ is the parallel propagator for this connection and $\widetilde R^A{}_{Bcd}$ its curvature tensor (see~\cite{Flanagan2016} and Secs.~III.A-C of Paper~I for further discussion).
By expanding the separation vector $\xi^{a''}$ in this integral in terms of the moments of the curve deviation observable and the flat-space separation vector $\xi_{\rm flat}^{a''}$, it seems plausible that one could write this observable in terms of moments as well.
Similarly, Eqs.~(4.126), (4.127), and~(4.180) of~\cite{GrantThesis} [a special case of the latter of which appeared as Eq.~(B3) of Paper~II] provide evidence that the dependence of the holonomy on initial separation, relative velocity, and the acceleration can be related to one another.
However, this is somewhat more complicated than the discussion in this paper, and so we will leave such an analysis to future work.

Using the framework of moments for the curve deviation in Sec.~\ref{sec:nonlinear}, it was straightforward to derive the results of Paper~III for the values of the curve deviation in terms of moments of the Bondi news, which only hold at $O(1/r)$.
However, these expressions provide a path forward to compute the curve deviation at $O(1/r^2)$ that is far simpler than starting with the original expressions.
This motivates finding expressions for the holonomy observables in Paper~I in terms of moments, as interesting differences between the various holonomies considered in Paper~I are expected to arise at this order, since the expressions at this order are no longer given by Paper~I's leading-order-in-curvature results.

There is a particular reason for interest in the $O(1/r^2)$ expressions for the holonomy observables considered in Paper~I.
The connections which were considered in Paper~I were inspired by transport laws which related linear and angular momentum at different points~\cite{Flanagan2014}.
Paper~I, building on~\cite{Flanagan2016}, defined a four-parameter family of connections $\vkappa \nabla_a$, characterized by a quadruple of scalar constants collectively referred to as $\varkappa$.
The transport law for linear and angular momentum in flat spacetime~\cite{Flanagan2014} corresponds to $\varkappa = (0, 0, 0, 0)$, and a transport law which is related both to the Mathisson-Papapetrou equations for spinning test body motion and to the transport equations for Killing vectors (see, for example,~\cite{Dixon1979, Harte2014}) is given by $\varkappa = (1/2, 0, 0, 0)$.
However, a new transport law, given by $\varkappa = (-1/4, 0, 0, 0)$\footnote{It was later argued in Paper~I (see the proof at the end of Sec.~4.1.3 of~\cite{GrantThesis}) that the more natural choice is $\varkappa = (-1/4, 1/2, 0, 0)$; the discrepancy is due to the fact that the last three parameters in $\varkappa$ constrain how the connection depends on the Ricci tensor, and~\cite{Flanagan2016} was concerned with either the vacuum Kerr spacetime or with asymptotically flat spacetimes, where the Ricci tensor falls off more rapidly.}, was singled out in~\cite{Flanagan2016} as being interesting in asymptotically flat spacetimes.
This was due to the following properties:
\begin{itemize}

\item it was the transport law obeyed by the closed, conformal Killing-Yano tensor in the Kerr spacetime (which has other connections to angular momentum, see for example~\cite{Floyd1974} and Sec.~6.3 of~\cite{Penrose1988}); and

\item it was shown that there existed an asymptotic solution to this transport law which was independent of the path taken, for stationary, asymptotically flat spacetimes.

\end{itemize}
The path-independence of this transport law for this value of $\varkappa$ implies that the holonomy with respect to this connection is just given by the identity, and so $\vkappa \Omega^A{}_B$ vanishes, but \emph{only} for this value of $\varkappa$, in these stationary spacetimes.
However, the results of Paper~I indicate that there is essentially no difference between the holonomies at $O(1/r)$, and so determining the relationship with the results of~\cite{Flanagan2016} would require an investigation of the holonomy to at least $O(1/r^2)$.

Finally, this series of papers has had as its primary focus the three observables defined in Paper~I: the curve deviation, the angular momentum holonomy, and finally an observable defined in terms of spinning test particles.
However, there are a plethora of other observables which have been considered in the literature, such as the Sagnac-interferometer observable which served as the original definition of the spin memory~\cite{Pasterski2015a}, or the orientation of a gyroscope relative to a fixed, fiducial orientation~\cite{Seraj2021b, Seraj2022a}.
Much like the analysis in this paper of the proper time observable of~\cite{Strominger2014b}, it would be valuable to perform nonlinear analyses of these observables as well.
In particular, there may be possible connections between these observables and those of Paper~I, and (as mentioned above) a nonlinear analysis may be helpful in determining how they behave at subleading orders.

\section{Acknowledgments}

The author thanks Abraham Harte and David Nichols for reading an early version of this paper and providing feedback.
The author also acknowledges the support of the Royal Society under grant number {RF\textbackslash ERE\textbackslash 221005}.
Initial work in this paper was supported by NSF Grant No. PHY-1707800 to Cornell University.

\appendix

\section{Coincidence limits of symmetrized derivatives} \label{app:symmetrized}

In order to derive the form of Taylor's theorem in Eq.~\eqref{eqn:scalar_taylor}, we need to show that symmetrized derivatives of Synge's world function take the following form:
\begin{equation} \label{eqn:symm_synge}
  [\sigma^a{}_{(b_1' \cdots b_n')}]_{x' \to x} = \begin{cases}
    -\delta^a{}_{b_1} & n = 1 \\
    0 & n \neq 1
  \end{cases}.
\end{equation}
Moreover, in order to derive the form in Eq.~\eqref{eqn:general_taylor}, we need to show that symmetrized derivatives of the parallel propagator vanish:
\begin{equation} \label{eqn:symm_parallel}
  [\nabla_{(c_1'} \cdots \nabla_{c_n')} g^{a'}{}_b]_{x' \to x} = \begin{cases}
    \delta^a{}_b & n = 0 \\
    0 & n \neq 0
  \end{cases}.
\end{equation}
In this appendix, we show that both of these results hold.

Starting with Eq.~\eqref{eqn:synge_geodesic}, we have that
\begin{equation} \label{eqn:synge_derivatives}
  \sigma^a{}_{(b_1' \cdots b_n')} = \sum_{k = 0}^n \binom{n}{k} \sigma^c{}_{(b_1' \cdots b_k'} \sigma^a{}_{b_{k + 1}' \cdots b_n')c}.
\end{equation}
In this sum, if $k = 0$, then the coincidence limit vanishes.
If $k = 1$, then we have by Synge's rule that
\begin{equation}
  \begin{split}
    [\sigma^c{}_{(b_1'} \sigma^a{}_{b_2' \cdots b_n')c}]_{x' \to x} = &-\nabla_{(b_1} [\sigma^a{}_{b_2' \cdots b_n')}]_{x' \to x} \\
    &+ [\sigma^a{}_{(b_1 \cdots b_n')}]_{x' \to x},
  \end{split}
\end{equation}
which involves the $n - 1$ case in the first term, and the same as the left-hand side in the second.
If $k = n$, then we find the same expression as the left-hand side of Eq.~\eqref{eqn:synge_derivatives} in the coincidence limit; note that if $n = 1$, this is not a distinct case from the above.
As such, the general expression for $n \geq 2$ is given by
\begin{equation}
  \begin{split}
    [\sigma^a{}_{(b_1' \cdots b_n')}]_{x' \to x} &= \nabla_{(b_1} [\sigma^a{}_{b_2' \cdots b_n')}]_{x' \to x} \\
    &\hspace{1.1em}- \sum_{k = 2}^{n - 1} \binom{n}{k} \frac{[\sigma^c{}_{(b_1' \cdots b_k'} \sigma^a{}_{b_{k + 1}' \cdots b_n')c}]_{x' \to x}}{n}.
  \end{split}
\end{equation}
In the case $n = 2$, the second term on the right-hand side vanishes as the sum is empty, whereas the first term vanishes because $[\sigma^a{}_{b'}]_{x' \to x}$ is a constant.
Because the terms on the right-hand side are recursively given in terms of the $n - 1$ and lower cases, this means that the left-hand side must vanish when $n \geq 2$ as well, by induction.
As such, including the cases $n = 0$ [coming from Eq.~\eqref{eqn:sep_coincidence}] and $n = 1$ [coming from Eq.~\eqref{eqn:synge_coincidence}], we find Eq.~\eqref{eqn:symm_synge}.

To derive Eq.~\eqref{eqn:symm_parallel}, we start with Eq.~\eqref{eqn:parallel_def}, which implies that
\begin{equation}
  \nabla_{b_1'} \cdots \nabla_{b_n'} (\sigma^{c'} \nabla_{c'} g^{a'}{}_a) = 0.
\end{equation}
Upon expanding the left-hand side and symmetrizing over $b_1' \cdots b_n'$, we therefore find that
\begin{equation}
  0 = \sum_{k = 0}^n \binom{n}{k} \sigma^c{}_{(b_1' \cdots b_k'} \nabla_{b_{k + 1}'} \cdots \nabla_{b_n')} \nabla_c g^{a'}{}_a.
\end{equation}
Taking a coincidence limit of this expression, by Eq.~\eqref{eqn:symm_synge}, the only non-zero term in the sum comes from $k = 1$, and that non-zero contribution is given by a symmetrized derivative of the parallel propagator.
Combining this result for $n > 0$ with the fact that $g^a{}_b = \delta^a{}_b$, we find that Eq.~\eqref{eqn:symm_parallel} holds.

\bibliographystyle{utphys}
\bibliography{Refs}

\end{document}